\documentclass{article}
\usepackage{amssymb}

\newtheorem{lemma}{Lemma}[section]
\newtheorem{theorem}[lemma]{Theorem}

\title{\Large\bf Point configurations, Cremona transformations and the 
elliptic difference Painlev\'e equation}

\author{\normalsize 
K.~Kajiwara${}^{1}$, T.~Masuda${}^{2}$, 
M.~Noumi${}^{2}$, Y.~Ohta${}^{2}$
and Y.~Yamada${}^{2}$\\
{\small ${}^{1}$\ Graduate School of Mathematics, Kyushu University,}\\[-4pt]
{\small Hakozaki, Fukuoka 812-8512, Japan}\\[-2pt]
{\small ${}^{2}$ Department of Mathematics, Kobe University,}\\[-4pt]
{\small Rokko, Kobe 657-8501, Japan}
}
\date{}

\newcommand{\bm}{\mbox{\boldmath$m$}}
\newcommand{\vep}{\varepsilon}

\newcommand{\br}[1]{{\langle{#1}\rangle}}
\newcommand{\ipr}[2]{({#1}\,|\,{#2})}

\newcommand{\iso}{\stackrel{\sim}{\to}}
\newcommand{\dfrac}[2]{{\displaystyle\frac{#1}{#2}}}
\newcommand{\dsum}[2]{{\displaystyle\sum_{#1}^{#2}}}
\newcommand{\dprod}[2]{{\displaystyle\prod_{#1}^{#2}}}

\newcommand{\binom}[2]{{#1\choose #2}}
\newcommand{\eqref}[1]{$(${\rm\ref{#1}}$)$}

\newcommand{\re}{{\mbox{\scriptsize Re}}}
\newcommand{\comment}[1]{}
\newcommand{\ratto}{
\begin{picture}(22,4)
\put(9.2,0){$\to$}
\multiput(3,0)(2,0){3}{$\cdot$}
\end{picture}
}

\newcommand{\twonodes}[2]{
\begin{picture}(40,10)
\put(10,2){\circle{4}}
\put(30,2){\circle{4}}
\put(2,0){\small$#1$}
\put(34,0){\small$#2$}
\end{picture}
}
\newcommand{\twonodesB}[2]{
\begin{picture}(40,10)
\put(10,2){\circle{4}}
\put(30,2){\circle{4}}
\put(12,2){\line(1,0){16}}
\put(2,0){\small$#1$}
\put(34,0){\small$#2$}
\end{picture}
}
\begin{document}

\maketitle

\begin{quote}{\small 
{\bf Abstract.}\ 
A theoretical foundation for a generalization of the elliptic 
difference Painlev\'e equation to higher dimensions is provided 
in the framework of birational Weyl group action on the space 
of point configurations in general position in a projective space.  
By introducing an elliptic parametrization of point configurations, 
a realization of the Weyl group is proposed as a group of Cremona transformations containing elliptic functions in the coefficients. 
For this elliptic Cremona system, a theory of $\tau$-functions is 
developed to translate it into a system of bilinear equations of 
Hirota-Miwa type for the $\tau$-functions on the lattice. 
Application of this approach is also discussed 
to the elliptic difference Painlev\'e equation. 
%\newline
%{\bf Keywords:}\ \ discrete Painlev\'e equation, 
%Cremona transformation, configuration space, 
%elliptic function
%\newline
%{\bf 2000 Mathematics Subject Classification:}\ 
%39A20; 14E07,14N20,14H52,33E17
}
\end{quote}

% dans la langue de l'article
\section{Introduction} 
\label{introduction}
%etc...

The main purpose of this paper is to provide a theoretical 
foundation for a generalization 
of the elliptic difference Painlev\'e equation 
to higher dimensions 
in the framework of birational Weyl group actions on the spaces of 
point configurations in general position 
in projective spaces. 

\par\medskip
Since the pioneering work of Grammaticos, Ramani, Papageorgiou and 
Hietarinta 
\cite{GRPH}, discrete Painlev\'e equations have been studied 
from various viewpoints.  A large class of second order 
discrete Painlev\'e equations, as well as their generalizations, 
has been discovered through the studies of singularity 
confinement property, % \cite{RGP}, 
bilinear equations, %\cite{RGS}, 
affine Weyl group symmetries %\cite{NY} 
and spaces of initial conditions %\cite{Sakai}.
(see \cite{RGP},\cite{RGS},\cite{NY},\cite{Sakai},$\ldots$). 
For historical aspects of discrete Painlev\'e equations, 
we refer the reader to the review of Grammaticos-Ramani \cite{GR}. 

Among many others, we mention here 
the geometric approach proposed by Sakai \cite{Sakai} 
for a class of discrete Painlev\'e equations arising 
from rational surfaces. 
Each equation in this class 
is defined by the group of Cremona transformations on a 
certain family of surfaces obtained from the projective plane 
$\mathbb{P}^2(\mathbb{C})$ by blowing-up. 
According to the types of rational surfaces, 
those discrete Painlev\'e equations are classified in 
terms of affine root systems. Also, their symmetries are 
described by means of affine Weyl groups. 
The {\em elliptic difference Painlev\'e equation}, 
which is regarded as 
the master equation for all discrete Painlev\'e 
equations of this class, 
is a discrete dynamical system 
defined on a family of surfaces parametrized by 
the 9-point configurations in general position in 
$\mathbb{P}^2(\mathbb{C})$; 
the corresponding group of Cremona transformations 
is the affine Weyl group of type $E^{(1)}_8$. 
As we have shown in \cite{KMNOY1}, 
this system of difference equations 
can be transformed into the eight-parameter discrete 
Painlev\'e equation of 
Ohta-Ramani-Grammaticos \cite{ORG}, 
constructed from a completely different viewpoint 
of bilinear equations for the $\tau$-functions on the 
$E_8$ lattice.
It is also known by \cite{KMNOY1} that 
the elliptic difference Painlev\'e equation has special 
Riccati type solutions obtained by linearization 
to the {\em elliptic difference hypergeometric equation}.
This gives a new perspective of nonlinear special 
functions to the elliptic hypergeometric functions 
which have been studied for instance by Frenkel-Turaev \cite{FT}
in the context of elliptic 6-$j$ symbols and 
by Spiridonov-Zhedanov \cite{SZ} 
in the theory of biorthogonal rational functions 
on elliptic grids. 

\par\medskip
Generalizing the geometric approach to the elliptic 
difference Painlev\'e equations, 
in this paper we investigate the configuration space 
$\mathbb{X}_{m,n}$ of $n$ points 
$p_1,\ldots,p_n$ in general position in the 
projective space $\mathbb{P}^{m-1}(\mathbb{C})$. 
It is well-known \cite{DO} that 
the Weyl group $W_{m,n}$ 
associated with the tree $T_{2,m,n-m}$ 
can be realized as a group of birational transformations 
on the configuration space $\mathbb{X}_{m,n}$. 
Through the $W_{m,n}$-equivariant projection 
$\mathbb{X}_{m,n+1}\to\mathbb{X}_{m,n}$ that maps 
$[p_1,\ldots,p_n,q]$ to $[p_1,\ldots,p_n]$, 
from the birational action of $W_{m,n}$ on $\mathbb{X}_{m,n+1}$ 
we obtain a realization of the Weyl group $W_{m,n}$ as a 
group of Cremona transformations on $q\in\mathbb{P}^{m-1}(\mathbb{C})$
parameterized by the configuration space $\mathbb{X}_{m,n}$. 
Note that in the case when $(m,n)=(3,9), (4,8)$ or $(6,9)$, the 
Weyl group $W_{m,n}$ is the affine Weyl group of type 
$E^{(1)}_8$, $E^{(1)}_7$ or $E^{(1)}_8$, respectively; 
%the affine Weyl group $W_{m,n}=W(E^{(1)}_l)$ then decomposes 
this group $W_{m,n}=W(E^{(1)}_l)$ decomposes 
into the semidirect product of the root lattice $Q(E_l)$ 
and the finite Weyl group $W(E_l)$. 
In each of the three cases, 
through the birational action of $W_{m,n}$ on $\mathbb{X}_{m,n+1}$, 
the lattice part of the affine Weyl group 
provides a {\em discrete Painlev\'e system} 
on $\mathbb{P}^{m-1}(\mathbb{C})$ 
with parameter space $\mathbb{X}_{m,n}$. 
The discrete Painlev\'e system of type $(3,9)$ thus obtained 
contains the three 
discrete Painlev\'e equations, elliptic, trigonometric and rational, 
with $W(E^{(1)}_8)$ symmetry in Sakai's table. 

In this framework of configuration spaces, in Section \ref{linearization} 
we construct a $W_{m,n}$-equivariant meromorphic mapping 
$\varphi_{m,n}: \mathfrak{h}_{m,n}\ratto \mathbb{X}_{m,n}$
by means of elliptic functions, where $\mathfrak{h}_{m,n}$ denotes 
the Cartan subalgebra of the Kac-Moody Lie algebra associated with 
the tree $T_{2,m,n-m}$. 
If we regard the birational $W_{m,n}$-action on $\mathbb{X}_{m,n}$ 
as a system of functional equations for the coordinate functions, 
a `canonical' elliptic solution is provided by the meromorphic mapping 
$\varphi_{m,n}$. 
Its image also specifies a $W_{m,n}$-stable class of $n$-point 
configurations in $\mathbb{P}^{m-1}(\mathbb{C})$ in which 
the $n$ points are on an elliptic curve. 
By restricting the point configurations to this class, 
from the birational Weyl group action of $W_{m,n}$ on 
$\mathbb{X}_{m,n+1}$ 
we obtain a realization of $W_{m,n}$ as a group of 
Cremona transformations on $\mathbb{P}^{m-1}(\mathbb{C})$ 
parametrized by elliptic functions, 
which we call the {\em elliptic Cremona system} of type $(m,n)$. 
In Section \ref{taufunctions} we develop a theory of $\tau$-functions 
for this elliptic Cremona system of type $(m,n)$, and 
show that it is translated into a system of bilinear equations 
of Hirota-Miwa type for the $\tau$-functions on the lattice. 
After that we reconsider the case of the elliptic difference 
Painlev\'e system of type $(3,9)$ in the scope of the general 
setting of this paper. 
There we give explicit description for some of 
the discrete time evolutions, 
in terms of homogeneous coordinates 
in Section \ref{ellipticdifference}, 
and in the language of geometry of plane curves in Section 
\ref{planecurves}. 
%This part is an appendix to our previous paper \cite{KMNOY1}. 
\par\medskip
The $\tau$-function approach developed in this paper 
can be applied effectively to the study of special 
hypergeometric solutions of the elliptic 
Painlev\'e equation and its degenerations. 
Also, it is an important problem to {\em complete} the 
framework of $\mathbb{X}_{m,n}$ of point configurations in 
general position, so that it should contain all reasonable 
degenerate configurations as in Sakai's table. 
These subjects will be investigated in our subsequent papers. 

%\tableofcontents

\section{Point configurations and Cremona transformations}
\label{configurations}

Let $\mathbb{X}_{m,n}$ be the configuration space 
of $n$ points in general position 
in $\mathbb{P}^{m-1}(\mathbb{C})$ ($n>m>1$). 
We say that an $n$-tuple of points $(p_1,\ldots,p_n)$ in 
$\mathbb{P}^{m-1}(\mathbb{C})$ is 
{\em in general position}  
if $p_1,\ldots,p_n$ are mutually distinct, and 
$\#(H\cap\{p_1,\ldots,p_n\})<m$ 
for any hyperplane $H$ in $\mathbb{P}^{m-1}(\mathbb{C})$. 
We denote 
by $[p_1,\ldots,p_n]$ 
the corresponding {\em configuration}, 
namely, the equivalence class of $(p_1,\ldots,p_n)$ under the 
diagonal $PGL_m(\mathbb{C})$-action. 
By fixing a system of homogeneous coordinates for 
$\mathbb{P}^{m-1}(\mathbb{C})$, 
the configuration space $\mathbb{X}_{m,n}$ may be identified 
with the double coset space 
\begin{equation}\label{eq:GLMT}
\mathbb{X}_{m,n}=GL_m(\mathbb{C})\backslash 
\mbox{Mat}^\ast_{m,n}(\mathbb{C})\slash T_n, 
\end{equation}
where $\mbox{Mat}_{m,n}^\ast(\mathbb{C})$ stands for the space 
of all $m\times n$ complex matrices whose $m\times m$ minor determinants 
are all nonzero, 
and $T_n=(\mathbb{C}^\ast)^n$ for the diagonal subgroup of 
$GL_n(\mathbb{C})$. 
The configuration space $\mathbb{X}_{m,n}$ has the structure of 
an affine algebraic variety, isomorphic to a Zariski 
open subset of $\mathbb{C}^{(m-1)(n-m-1)}$ 
(see \cite{Yoshida}, for instance). 
Also, it is known \cite{Coble}, \cite{DO} that the Weyl group 
associated with the tree 
\begin{equation}
%\fbox{
\unitlength=1pt
\begin{picture}(205,25)(-55,-2)
\put(0,0){\circle{4}}\put(2,0){\line(1,0){16}}
\put(20,0){\circle{4}}\put(22,0){\line(1,0){8}}
\multiput(30,0)(4,0){6}{\line(1,0){1}}
\put(50,0){\line(1,0){8}}
\put(60,0){\circle{4}}
\put(62,0){\line(1,0){16}}
\put(80,0){\circle{4}}
\put(82,0){\line(1,0){8}}
\multiput(90,0)(4,0){6}{\line(1,0){1}}
\put(110,0){\line(1,0){8}}
\put(120,0){\circle{4}}
\put(122,0){\line(1,0){16}}
\put(140,0){\circle{4}}
\put(60,2){\line(0,1){14}}
\put(60,18){\circle{4}}
\put(-2,-8){\small$\alpha_1$}
\put(18,-8){\small$\alpha_2$}
\put(58,-8){\small$\alpha_m$}
\put(78,-8){\small$\alpha_{m+1}$}
\put(136,-8){\small$\alpha_{n-1}$}
\put(64,18){\small$\alpha_0$}
\put(-75,0){$T_{2,m,n-m}:$}
\end{picture}
%}
\vspace{4pt}
\end{equation}
acts birationally on $\mathbb{X}_{m,n}$.
This Weyl group $W_{m,n}=W(T_{2,m,n-m})$ is generated 
by the {\em simple reflections} 
$s_0,s_1,\ldots,s_{n-1}$ 
with the following fundamental relations. 
\begin{equation}
W_{m,n}=\langle s_0,s_1,\ldots,s_{n-1}\rangle:
\qquad\qquad
\renewcommand{\arraystretch}{0.8}
\begin{array}{cccccc}
s_i^2=1 &\quad&{\alpha_i\quad\ \alpha_j}\\
s_i s_j=s_j s_i &&\twonodes{}{}\\
s_i s_j s_i=s_j s_i s_j 
&&\twonodesB{}{}
\end{array}
\end{equation}
As we will recall below,  $W_{m,n}$ is realized as a group of 
birational transformations of $\mathbb{X}_{m,n}$ by 
the standard Cremona transformations 
with respect to $m$ points among $p_1,\ldots,p_n$. 

Given a set of $m$ points $p_1,\ldots,p_m$ in general position, 
choose a system of homogeneous coordinates $x=(x_1,\ldots,x_m)$ 
such that 
\begin{equation}
p_1=(1:0:\ldots:0),\ \ 
p_2=(0:1:\ldots:0),\ \ \ldots,\ \ 
p_m=(0:\ldots:0:1).
\end{equation}
Then the {\em standard Cremona transformation} 
with respect to $(p_1,\ldots,p_m)$ 
is the birational transformation $p\to \widetilde{p}$ 
of $\mathbb{P}^{m-1}(\mathbb{C})$ 
defined by 
%$\widetilde{p}=\left(\dfrac{1}{x_1}:\ldots:\dfrac{1}{x_m}\right)$ 
$\widetilde{p}=\left(x_1^{-1}:\ldots:x_m^{-1}\right)$ 
for any $p=(x_1:\ldots:x_m)$ with $x_i\ne 0$ ($i=1,\ldots,m$).
Note that 
this transformation depends on the choice of homogeneous coordinates, 
and is determined only up to the action of $(\mathbb{C}^\ast)^m$. 
The birational (right) action of 
$W_{m,n}$ on $\mathbb{X}_{m,n}$ is then defined as follows. 
Firstly, 
the symmetric group $\mathfrak{S}_n$ acts on $\mathbb{X}_{m,n}$ 
by the permutation of $n$ points: 
\begin{equation} 
[p_1,\ldots,p_n].\sigma=[p_{\sigma(1)},\ldots,p_{\sigma(n)}]
\qquad(\sigma\in\mathfrak{S}_n).
\end{equation}
The adjacent transpositions $s_j=(j,j+1)$ ($j=1,\ldots,n-1$) 
provide the simple reflections attached to the subdiagram of type 
$A_{n-1}$ in $T_{2,m,n-m}$. 
The remaining simple reflection $s_0$ is given 
by the (well-defined) birational transformation  
\begin{equation}
[p_1,\ldots,p_n].s_0
=[p_1,\ldots,p_m,\widetilde{p}_{m+1},\ldots,\widetilde{p}_n],
\end{equation}
in terms of the standard Cremona transformation 
$p\to \widetilde{p}$ with respect to the first $m$ 
points $(p_1,\ldots,p_m)$. 
These birational 
transformations $s_0, s_1,\ldots,s_{n-1}$ in fact satisfy the 
fundamental relations for the simple reflections of 
$W_{m,n}$. 
We also remark that, 
for each subset $\{j_1,\ldots,j_m\}\subset\{1,\ldots,n\}$ 
of mutually distinct $m$ indices, 
the standard Cremona transformation 
with respect to $(p_{j_1},\ldots,p_{j_m})$ 
is determined as $\mbox{cr}_{j_1,\ldots,j_m}=\sigma s_0 \sigma^{-1}
\in W_{m,n}$ 
by a permutation $\sigma\in\mathfrak{S}_n$ such 
that $\sigma(a)=j_a$ for $a=1,\ldots,m$. 

The right birational action of $W_{m,n}$ on $\mathbb{X}_{m,n}$ 
induces a left action of $W_{m,n}$ on the field 
$\mathcal{K}(\mathbb{X}_{m,n})$ of rational functions on $\mathbb{X}_{m,n}$ 
as a group of automorphisms: 
For each $\varphi\in \mathcal{K}(\mathbb{X}_{m,n})$ and $w\in W_{m,n}$, 
we define $w(\varphi)\in \mathcal{K}(\mathbb{X}_{m,n})$ by 
\begin{equation} 
w(\varphi)([p_1,\ldots,p_n])=\varphi([p_1,\ldots,p_n].w)
\end{equation}
for any generic $[p_1,\ldots,p_n]\in\mathbb{X}_{m,n}$. 
Let us consider the set $\mathcal{U}_{m,n}$ of all matrices 
$U\in\mbox{Mat}_{m,n}(\mathbb{C})^\ast$ of the form
\begin{equation}
U=\left(\matrix{
1 & \ldots & 0 & 0 & 1 & u_{1,m+2} & \ldots & u_{1,n} \cr
\vdots&\ddots&\vdots&\vdots&\vdots&\vdots& & \vdots\cr
0 & \ldots & 1 & 0 & 1 & u_{m-1,m+2} & \ldots & u_{m-1,n} \cr
0 & \ldots & 0 & 1 & 1 & 1 & \ldots &1 
}\right). 
\end{equation} 
It is easily shown that each $(GL_m(\mathbb{C}),T_n)$-orbit in 
$\mbox{Mat}_{m,n}^\ast(\mathbb{C})$ intersects with 
$\mathcal{U}_{m,n}$ at one point. 
By using this transversal 
$\mathcal{U}_{m,n}\iso GL_m(\mathbb{C})\backslash
\mbox{Mat}_{m,n}(\mathbb{C})^\ast\slash T_n$, 
we identify 
$\mathbb{X}_{m,n}$ with a Zariski open subset of 
the affine space $\mathbb{C}^{(m-1)(n-m-1)}$ with 
canonical coordinates $u=(u_{i,j})_{1\le i\le m-1; m+2\le j\le n}$. 
Through the isomorphism 
$\mathcal{K}(\mathbb{X}_{m,n})\iso\mathbb{C}(u)$, 
the action of $W_{m,n}$ 
on $\mathcal{K}(\mathbb{X}_{m,n})$ can be described explicitly 
in terms of the $u$ variables. 
The following table shows how the simple reflections  
$s_k$ ($k=0,1,\ldots,n-1$) transform the coordinates  
$u_{i,j}$ ($i=1,\ldots,m-1; j=m+2,\ldots,n$): 
\begin{equation}\label{eq:Wmns}
\arraycolsep=2pt
\begin{array}{lrlll}
k=0:\qquad& s_0(u_{ij})&=\dfrac{1}{u_{ij}},
\\
k=1,\ldots,m-2: & s_k(u_{ij})&=u_{s_k(i),j},
\\[6pt]
k=m-1: & s_{m-1}(u_{ij}) &=
\left\{
\begin{array}{ll}
\dfrac{u_{ij}}{u_{m-1,j}}\quad& (i=1,\ldots,m-2),\\[8pt]
\dfrac{1}{u_{m-1,j}} & (i=m-1),
\end{array}
\right.
\\[20pt]
k=m: & s_{m}(u_{ij})&=1-u_{ij},
\\[6pt]
k=m+1: & s_{m+1}(u_{ij}) &=
\left\{
\begin{array}{ll}
\dfrac{1}{u_{i,m+2}} & (j=m+2),\\[8pt]
\dfrac{u_{ij}}{u_{i,m+2}}\quad& (j=m+3,\ldots,n),
\end{array}
\right.
\\
k=m+2,\ldots,n-1: & 
s_k(u_{ij})&=u_{i,s_k(j)},
\end{array}
\end{equation}
where $s_k(i)$ stands for the action of 
the adjacent transposition 
$(k,k+1)$ on the index $i\in\{1,\ldots,n\}$.
From this representation, for each $w\in W_{m,n}$ we obtain 
a family of rational functions 
\begin{equation}\label{eq:Wmn}
w(u_{i,j})=S^{w}_{i,j}(u)\qquad(i=1,\ldots,m-1; j=m+2,\ldots,n)
\end{equation}
in the $u$ variables; 
these functions satisfy the consistency relations 
\begin{equation}
S^{1}_{i,j}(u)=u_{ij},
\qquad
S^{ww'}_{i,j}(u)=S_{i,j}^{w'}(S^w(u))
\end{equation}
for any $i,j$ and $w,w'\in W_{m,n}$. 

If we regard the $u$ variables as dependent variables (unknown functions), 
\eqref{eq:Wmn} or equivalently \eqref{eq:Wmns} can be regarded as 
a system of functional equations for $u_{ij}$. 
Typically, we take a vector space $V$ with canonical 
coordinates $t=(t_1,\ldots,t_N)$, assuming that 
$W_{m,n}$ acts linearly 
on $V$ (from the right). 
If we regard the coordinate functions 
$t_j$ of $V$ as the independent variables, 
then a solution of the system \eqref{eq:Wmn} is nothing but 
a $W_{m,n}$-equivariant mapping $\varphi: V\to\mathbb{X}_{m,n}$. 
In Section \ref{linearization}, we construct an elliptic solution of the 
system \eqref{eq:Wmn} in this sense, with $V$ being 
the Cartan subalgebra $\mathfrak{h}_{m,n}$ 
of the Kac-Moody Lie algebra associated with $T_{2,m,n-m}$. 

\section{Tracing the Cremona transformations}
\label{Cremona}

Given a generic configuration $[p_1,\ldots,p_n]$ of $n$ points,  
let us ask how a general point of $\mathbb{P}^{m-1}(\mathbb{C})$, 
as well as the configuration itself, 
is transformed by a successive application 
of standard Cremona transformations. 
In what follows, by a {\em Cremona transformation} we mean 
a birational transformation of $\mathbb{P}^{m-1}(\mathbb{C})$ 
obtained by a successive application of standard Cremona 
transformations. 

\par\medskip 
We now consider the relative situation 
with respect to the projection 
$\pi: \mathbb{X}_{m,n+1} \to\mathbb{X}_{m,n}$
that maps $[p_1,\ldots,p_n,p_{n+1}]$ to 
$[p_1,\ldots,p_n]$. 
This projection is $W_{m,n}$-equivariant relative 
to the inclusion $W_{m,n}\subset W_{m,n+1}$ of 
Weyl groups. We regard $\mathbb{X}_{m,n}$ as 
the parameter space for Cremona transformations 
belonging to $W_{m,n}$, 
and the last point $q=p_{n+1}$ as the general point in 
$\mathbb{P}^{m-1}(\mathbb{C})$ that should be 
transformed by such Cremona transformations. 
(This formulation has been used by \cite{MY} in the case $(m,n)=(3,9)$.)
Then, our question is how to describe 
$[p_1,\ldots,p_n,q].w$ 
for each $[p_1,\ldots,p_n,q]\in \mathbb{X}_{m,n+1}$ and 
$w\in W_{m,n}$. By using the coordinates 
\begin{equation}
\widetilde{U}=\left(\matrix{
1 & \ldots & 0 & 0 & 1 & u_{1,m+2} & \ldots & u_{1,n} & z_1\cr
\vdots&\ddots&\vdots&\vdots&\vdots&\vdots& & \vdots & \vdots\cr
0 & \ldots & 1 & 0 & 1 & u_{m-1,m+2} & \ldots & u_{m-1,n} & z_{m-1}\cr
0 & \ldots & 0 & 1 & 1 & 1 & \ldots &1 &1 
}\right)
\end{equation}
for $\mathcal{U}_{m,n+1}$, 
we parametrize the configurations 
$[p_1,\ldots,p_n]\in\mathbb{X}_{m,n}$ and 
the general points $q\in\mathbb{P}^{m-1}(\mathbb{C})$ 
as 
\begin{equation}\label{eq:pq}
\arraycolsep=2pt
\begin{array}{ll}
p_1=(1:0:\ldots:0),\ \ldots,\ \ p_{m}=(0:\ldots:0:1),\ \ 
p_{m+1}=(1:\ldots:1:1),\\
p_{j}=(u_{1,j}:\ldots:u_{m-1,j}:1)\ \ (j=m+2,\ldots,n),\quad
q=(z_1:\ldots:z_{m-1}:1). 
\end{array}
\end{equation}
Then for any $w\in W_{m,n}$ the configuration 
$[p_1,\ldots,p_n,q].w=[\widetilde{p}_1,\ldots,\widetilde{p}_n,\widetilde{q}]$ 
is given by
\begin{equation}
\arraycolsep=2pt
\begin{array}{rll}
\widetilde{p}_j&=p_j\qquad&(j=1,\ldots,m+1),\\
\widetilde{p}_{j}&=(w(u_{1,j}):\ldots:w(u_{m-1,j}):1)\qquad&(j=m+2,\ldots,n),\\
\widetilde{q}&=(w(z_1):\ldots:w(z_{m-1}):1). 
\end{array}
\end{equation}
In this sense, for each $w\in W_{m,n}$ 
the corresponding Cremona transformation 
of $q=p_{n+1}$ 
is determined as 
\begin{equation}
w(z_i)=R^{w}_{i}(u;z)\qquad(i=1,\ldots,m-1), 
\end{equation}
in terms of rational functions $R^{w}_{i}(u;z)$
in the variables $u=(u_{ij})_{1\le i\le m-1; m+2\le j\le n}$
and $z=(z_1,\ldots,z_{m-1})$. 
Note also that $R^{w}_i(u;z)$ satisfy
\begin{equation}
R^{1}_i(u;z)=z_i,\qquad
R^{ww'}_i(u;z)=R^{w'}_i(S^{w}(u);R^{w}(u;z))
\end{equation}
for any $i$ and $w,w'\in W_{m,n}.$ 
As we will see below, 
these $R^{w}_i(u;z)$, regarded as rational functions in the variable $z=(z_1,\ldots,z_{m-1})$, have a characteristic property 
concerning their multiplicities of zero at $p_1,\ldots,p_n$. 

Consider a free $\mathbb{Z}$-module 
\begin{equation}
L_{m,n}=\mathbb{Z}e_0\oplus \mathbb{Z}e_1\oplus\cdots\oplus 
\mathbb{Z}e_n
\end{equation}
of rank $n+1$ with basis $\{e_0,e_1,\ldots,e_n\}$, 
and define a symmetric bilinear form 
$\ipr{}{} : L_{m,n}\times L_{m,n} \to \mathbb{Z}$ by 
\begin{equation}
\begin{array}{cc}
\ipr{e_0}{e_0}=-(m-2),\quad 
\ipr{e_j}{e_j}=1\quad(j=1,\ldots,n),\\
\ipr{e_i}{e_j}=0\qquad(i,j=0,1,\ldots,n;\ i\ne j). 
\end{array}
\end{equation}
This lattice $L_{m,n}$ admits a natural linear action 
of the Weyl group $W_{m,n}$ defined by
\begin{equation}
s_k.\Lambda=\Lambda-\ipr{h_k}{\Lambda}\,h_k
\qquad(\Lambda\in L_{m,n})
\end{equation} 
for each $k=0,1,\ldots,n-1$, where 
\begin{equation}
h_0=e_0-e_1-\cdots-e_m,\quad
h_k=e_k-e_{k+1}\quad(k=1,\ldots,n). 
\end{equation}
Note that $\ipr{h_j}{h_j}=2$ ($j=0,1,\ldots,n-1$) 
and that 
$(\ipr{h_i}{h_j})_{i,j=0}^{n-1}$ is the 
(generalized) Cartan matrix associated with the tree $T_{2,m,n-m}$. 
For each 
\begin{equation}
\Lambda=d e_0-\nu_1e_1-\cdots-\nu_ne_n\in L_{m,n} 
\qquad(d,\nu_1,\ldots,\nu_n\in\mathbb{Z}) 
\end{equation}
we denote by  
$L(\Lambda)$ the vector space 
over $\mathcal{K}(\mathbb{X}_{m,n})$ 
consisting of all homogeneous polynomials 
$f(x)\in\mathcal{K}(\mathbb{X}_{m,n})[x]$ 
of degree $d$ 
in $m$ variables $x=(x_1,\ldots,x_m)$ 
that have a zero of multiplicity $\ge \nu_j$ at each $p_j$ 
($j=1,\ldots,n$): 
\begin{equation}
\deg\,f(x)=d,\qquad
\mbox{ord}_{p_j}f(x)\ge \nu_j\quad(j=1,\ldots,n). 
\end{equation} 
Here we regard $x=(x_1,\ldots,x_m)$ as the homogeneous 
coordinate system for $\mathbb{P}^{m-1}(\mathcal{K}(\mathbb{X}_{m,n}))$
such that $(z_1:\ldots:z_{m-1}:1)=(a_1x_1:\cdots:a_mx_m)$
for some nonzero constants $a_i\in\mathcal{K}(\mathbb{X}_{m,n})$ 
($i=1,\ldots,m-1$). 
Then we have 
\begin{theorem}\label{thm:GF}
$(1)$ Let $M_{m,n}=W_{m,n}\{e_1,\ldots,e_n\}$ be the $W_{m,n}$-orbit 
of $\{e_1,\ldots,e_n\}$ in $L_{m,n}$. 
Then for any $\Lambda\in M_{m,n}$, one has 
$\dim_{\mathcal{K}(\mathbb{X}_{m,n})} L(\Lambda)=1$. 
\newline
$(2)$ Given any element $w\in W_{m,n}$, take nonzero polynomials
\begin{equation}
F_i(x)\in L(w.e_i),\quad G_i(x)\in L(ws_0.e_i)
\qquad(i=1,\ldots,m). 
\end{equation}
Then one has
\begin{equation}
(w(z_1):\ldots:w(z_{m-1}):1)
=\left(c_1\dfrac{G_1(x)}{F_1(x)}:\cdots: 
c_m\dfrac{G_m(x)}{F_m(x)}\right)
\end{equation} 
for some nonzero constants 
$c_i\in\mathcal{K}(\mathbb{X}_{m,n})$ $(i=1,\ldots,m)$. 
\end{theorem}
This theorem can be proved by decomposing each $w\in W_{m,n}$
into a product of simple reflections $w=s_{j_1}\cdots s_{j_p}$, 
and then by lifting each $s_{j_k}$ to the level of homogeneous 
coordinates.  We remark that there is no canonical way 
{\em a priori} to define the action of $s_j$ on homogeneous 
polynomials.  This is the reason why we cannot 
specify the choice of $F_i$ and $G_i$. 
We will return to this point later in Section \ref{taufunctions} 
in the context of $\tau$-functions for the elliptic Cremona system. 
In the case $(m,n)=(3,9)$, for any $\Lambda\in M_{m,n}$ a nontrivial 
element in $L(\Lambda)$ can be constructed as a certain interpolation 
determinant(see Section \ref{ellipticdifference}). 

\par\medskip
As we have seen above, the birational action of $W_{m,n}$ on 
$\mathbb{X}_{m,n+1}$ can be expressed in the form 
\begin{equation}\label{eq:SR} 
w(u_{i,j})=S^{w}_{i,j}(u),\qquad 
w(z_i)=R^{w}_i(u;z). 
\end{equation}
Formula \eqref{eq:SR} can be thought of as 
a system of functional equations for the dependent variables 
$z=(z_1,\ldots,z_{m-1})$ including the $u$ variables as 
parameters. 
Theorem \ref{thm:GF} then implies that such a system of equations 
can be expressed as 
\begin{equation}\label{eq:ECrmn}
w(z_i)=R^{w}_i(z)=\dfrac{c_i}{c_m}\dfrac{G^w_i(x)F^w_m(x)}{F^w_i(x)G^w_m(x)}
\qquad(i=1,\ldots,m-1)
\end{equation}
for each $w\in W_{m,n}$, 
by means of homogeneous polynomials $F^{w}_i\in L(w.e_i)$ and 
$G^w_i\in L(ws_0.e_i)$ that are characterized by the degree 
and the multiplicities of zero at $p_1,\ldots,p_n$. 
(The dependence on the $u$ variables is suppressed in this formula.) 
Note that any abelian subgroup of the Weyl group $W_{m,n}$ 
gives rise to a commuting family of birational transformations 
on $\mathbb{P}^{m-1}(\mathbb{C})$ parameterized by 
configurations of $n$ points. 
Such a birational dynamical system (for the $z$-variables)
could be called a {\em discrete Painlev\'e system} 
associated with point configurations in $\mathbb{P}^{m-1}(\mathbb{C})$. 

When the number 
$4-(m-2)(n-m-2)$ 
has the sign $+,0,$ or $-$, 
the root system associated with the tree $T_{2,m,n-m}$ is  
of finite type, of affine type, or of indefinite type, 
respectively, in the sense of \cite{Kac}. 
In particular there are three cases $(m,n)=(3,9), (4,8)$ and $(6,9)$ 
of affine type; the corresponding root systems are 
of type $E^{(1)}_8$, $E^{(1)}_7$ and $E^{(1)}_8$, respectively. 
In these three cases, 
the configuration of $n$ points can be parametrized generically 
by means of elliptic functions. 
Note also that the Weyl group 
$W_{m,n}=W(E^{(1)}_{l})$ is then expressed as the 
semidirect product of 
the root lattice $Q(E_l)$ of rank $l$ and 
the finite Weyl group $W(E_{l})$ acting on it. 
We call the discrete dynamical system arising from the 
lattice part of $W_{m,n}$ the {\em elliptic difference 
Painlev\'e equation} of type $(m,n)$. 

\section{Linearization of the $W_{m,n}$-action in terms of 
elliptic functions}
\label{linearization}

Let 
$\mathfrak{h}_{m,n}=L_{m,n}\otimes_{\mathbb{Z}}\mathbb{C}$ the 
complexification of the lattice $L_{m,n}$. 
In this section 
we construct a $W_{m,n}$-equivariant meromorphic mapping 
$\varphi_{m,n}:\mathfrak{h}_{m,n}\ratto\mathbb{X}_{m,n}$
by means of elliptic functions. 
This mapping specifies a class of configurations 
of $n$ points on an elliptic curve in 
$\mathbb{P}^{m-1}(\mathbb{C})$, which is preserved by 
the action of $W_{m,n}$. 
In order to simplify the presentation, we assume 
$m\ge 3$. In this case 
the symmetric bilinear form $\ipr{}{}: 
\mathfrak{h}_{m,n}\times \mathfrak{h}_{m,n}\to\mathbb{C}$ 
is nondegenerate, and 
$\mathfrak{h}_{m,n}$ is identified 
with the Cartan subalgebra of the Kac-Moody Lie algebra 
associated with the tree $T_{2,m,n-m}$.
\par\medskip
We define the linear functions $\vep_j$ ($j=0,1,\ldots,n$) 
and $\alpha_j$ ($j=0,1,\ldots,n-1$) on $\mathfrak{h}_{m,n}$ by
\begin{equation}
\vep_j(h)=\ipr{e_j}{h},\quad \alpha_j(h)=\ipr{h_j}{h}
\qquad(h\in\mathfrak{h}_{m,n}).
\end{equation} 
We regard $\vep=(\vep_0,\vep_1,\ldots,\vep_n): \mathfrak{h}_{m,n}\iso 
\mathbb{C}^{n+1}$ as the canonical 
coordinates for $\mathfrak{h}_{m,n}$. 
The linear functions $\alpha_0,\alpha_1,\ldots,\alpha_{n-1}$ are 
the {\em simple roots} of the root system associated with $T_{2,m,n-m}$. 
Note also that the dual space 
$\mathfrak{h}_{m,n}^\ast=
\mathbb{C}\vep_0\oplus
\mathbb{C}\vep_1\oplus\cdots\oplus
\mathbb{C}\vep_n$
has the induced symmetric bilinear form: 
$\ipr{\vep_i}{\vep_j}=\ipr{e_i}{e_j}$ 
for any $i,j\in\{0,1,\ldots,n\}$.  
The Weyl group $W_{m,n}$ acts on 
$\mathfrak{h}_{m,n}$ and $\mathfrak{h}_{m,n}^{\ast}$
in a standard way: For each $k=0,1,\ldots,n-1$, 
\begin{equation} 
s_k.h=h-\br{h,\alpha_k}h_k\quad(h\in\mathfrak{h}_{m,n}),
\quad
s_k.\lambda=\lambda-\br{h_k,\lambda}\alpha_k
\quad(\lambda\in\mathfrak{h}_{m,n}^\ast), 
\end{equation}
where $\br{\ ,\ }:\mathfrak{h}_{m,n}\times\mathfrak{h}_{m,n}^\ast
\to\mathbb{C}$  
is the canonical pairing. 
When we consider the right action 
of $W_{m,n}$ on 
$\mathfrak{h}_{m,n}$, 
we use the convention $h.w=w^{-1}.h$ for 
$h\in \mathfrak{h}_{m,n}$ and $w\in W_{m,n}$. 

We now fix a nonzero holomorphic function on $\mathbb{C}$, 
denoted by $[x]$, assuming that $[x]$ is odd 
($[-x]=-[x]$ for any $x\in\mathbb{C}$),  
and satisfies the {\em Riemann relation}:
\begin{equation}
[x+y][x-y][u+v][u-v]=
[x+u][x-u][y+v][y-v]-[x+v][x-v][y+u][y-u]
\end{equation} 
for any $x,y,u,v\in\mathbb{C}$.  
If this condition is satisfied, 
the set $\Omega=\{a\in\mathbb{C}\ |\ [a]=0\}$ 
of zeros of $[x]$ forms a $\mathbb{Z}$-submodule of $\mathbb{C}$,
and the function $[x]$ is quasi-periodic with 
respect to $\Omega$. 
There are three classes of such functions; 
elliptic, trigonometric and rational, according to 
the rank of $\Omega$: 
\par\smallskip\qquad
\begin{tabular}{lll}
Elliptic case:&\quad 
$[x]=c\,e^{ax^2}\sigma(x;\Omega)$\quad &($\Omega=\mathbb{Z}\omega_1\oplus
\mathbb{Z}\omega_2$),\\
Trigonometric case:&\quad 
$[x]=c\,e^{ax^2}\sin(\pi x/\omega_0)$\qquad &($\Omega=\mathbb{Z}\omega_0$),\\
Rational case:&\quad 
$[x]=c\,e^{ax^2} x\quad$ &($\Omega=\{0\}$). 
\end{tabular}
\par\smallskip\noindent
Here $\sigma(x;\Omega)$ denotes the Weierstrass sigma function 
associated with the period lattice $\Omega$. 
In the context of discrete Painlev\'e equations, these three 
cases correspond to the three types of 
difference equations (elliptic, trigonometric and rational). 
We will use this symbol $[x]$ whenever the three cases can be 
treated simultaneously.  

Taking constants $\lambda\in\mathbb{C}$ and
$\mu=(\mu_1,\ldots,\mu_m)\in\mathbb{C}^m$ 
such that $[\lambda]\ne 0$ and $[\mu_i-\mu_j]\ne0 $ ($1\le i<j\le m$), 
we define a holomorphic mapping 
$p_{\lambda,\mu}:\mathbb{C}\to\mathbb{P}^{m-1}(\mathbb{C})$ by 
\begin{equation}
\arraycolsep=2pt
\begin{array}{rl}
p_{\lambda,\mu}(t)&=
\left(\dfrac{[\lambda+\mu_1-t]}{[\lambda][\mu_1-t]}:\ldots
:\dfrac{[\lambda+\mu_m-t]}{[\lambda][\mu_m-t]}\right)
\\
&=
([\lambda+\mu_1-t]\dprod{k=2}{m}[\mu_k-t]:\ldots:
[\lambda+\mu_m-t]\dprod{k=1}{m-1}[\mu_k-t])
\end{array}
\end{equation}
for any $t\in\mathbb{C}$. 
Thanks to the quasi-periodicity of $[x]$, this mapping induces 
a holomorphic mapping 
$\overline{p}_{\lambda,\mu}: 
E_{\Omega}=\mathbb{C}/\Omega\to\mathbb{P}^{m-1}(\mathbb{C})$.  
We denote by $C_{\lambda,\mu}=\overline{p_{\lambda,\mu}(\mathbb{C})}
\subset\mathbb{P}^{m-1}(\mathbb{C})$ the curve obtained as the closure 
of the image of $p_{\lambda,\mu}$.  Note that 
this curve passes the $m$ coordinate origins in 
$\mathbb{P}^{m-1}(\mathbb{C})$; in fact we have
\begin{equation}
p_{\lambda,\mu}(\mu_1)=(1:0:\ldots:0),\quad\ldots,\quad 
p_{\lambda,\mu}(\mu_m)=(0:\ldots:0:1).
\end{equation}
We also remark that $C_{\lambda,\mu}$ is a smooth elliptic curve 
when $\mbox{rank}\,\Omega=2$, and a singular elliptic curve 
with a node (resp. a cusp) when $\mbox{rank}\,\Omega=1$
(resp. $\mbox{rank}\,\Omega=0$) at $(1:\ldots:1)$. 

We now consider a configuration of $n$ points on 
$C_{\lambda,\mu}\subset\mathbb{P}^{m-1}(\mathbb{C})$ 
defined as 
\begin{equation}
[p_1,\ldots,p_n]=[p_{\lambda,\mu}(\vep_1),\ldots,p_{\lambda,\mu}(\vep_n)]
\end{equation}
by the coordinates $\vep_j\in\mathbb{C}$ $(j=1,\ldots,n)$. 
Setting $\vep_0=\lambda+\mu_1+\cdots+\mu_m$, we assume that 
the parameters $\vep=(\vep_0,\vep_1,\ldots,\vep_n)$ are generic 
in the sense 
\begin{equation}
\begin{array}{ll}
[\vep_i-\vep_j]\ne0,\qquad&(1\le i<j\le n),\\{}
[\vep_0-\vep_{j_1}-\cdots-\vep_{j_m}]\ne 0 
\quad& (1\le j_1<\ldots<j_m\le n). 
\end{array}
\end{equation} 
Then by the Frobenius formula
\begin{equation}
\det\left(\dfrac{[\lambda+x_i-y_j]}{[\lambda][x_i-y_j]}\right)_{i,j=1}^m
=\dfrac{[\lambda+\dsum{k=1}{m}(x_k-y_k)]
\dprod{1\le i<j\le m}{}[x_j-x_i][y_i-y_j]
}{[\lambda]\dprod{1\le i,j\le m}{}[x_i-y_j]}
\end{equation}
for the function $[x]$, we see that the configuration $[p_1,\ldots,p_n]$ 
defined as above is in general position, 
and that its $u$ coordinates are given explicitly by 
\begin{equation}\label{eq:uinep}
\begin{array}{c}
u_{i,j}=u_{i,j}(\vep)=
\dfrac{[\alpha_0+\vep_{m,m+1}][\vep_{i,m+1}]}
{[\vep_{m,m+1}][\alpha_0+\vep_{i,m+1}]}
\dfrac{[\alpha_0+\vep_{i,j}][\vep_{m,j}]}
{[\vep_{i,j}][\alpha_0+\vep_{m,j}]}\\[8pt]
(i=1,\ldots,m-1; j=m+2,\ldots,n)
\end{array}
\end{equation}
where 
$\vep_{i,j}=\vep_{i}-\vep_{j}$ for $i,j\in\{1,\ldots,n\}$,
and $\alpha_0=\vep_0-\vep_1-\cdots-\vep_m$.  
Note that, by the passage to the double coset space $\mathbb{X}_{m,n}$, 
the dependence of the configuration 
on the parameters $\lambda$ and $\mu=(\mu_1,\ldots,\mu_m)$ 
has been confined in the parameter $\vep_0=\lambda+\mu_1+\cdots+\mu_m$.
Observe also that these functions $u_{i,j}(\vep)$ are $\Omega$-periodic 
in all the variables $\vep_j$ ($j=0,1,\ldots,n$). 

Under the identification of the parameters $\vep=(\vep_0,\vep_1,\ldots,\vep_n)$ 
with the canonical coordinates for $\mathfrak{h}_{m,n}$, the 
construction described above implies two meromorphic mappings  
\begin{equation}
\arraycolsep=2pt 
\begin{array}{lcll}
\varphi_{m,n}: &
\mathfrak{h}_{m,n}=
L_{m,n}\otimes_{\mathbb{Z}}\mathbb{C}&\ratto&\mathbb{X}_{m,n},
\quad\mbox{and}
\\[4pt]
\overline{\varphi}_{m,n}: &
E_{m,n}=L_{m,n}\otimes_{\mathbb{Z}}(\mathbb{C}/\Omega)&\ratto&\mathbb{X}_{m,n}. 
\end{array}
\end{equation}
Note that, 
when $\vep=(\vep_0,\vep_1,\ldots,\vep_n)\in\mathfrak{h}_{m,n}$ is 
generic, the corresponding configuration 
$\varphi_{m,n}(\vep)=[p_1,\ldots,p_n]\in\mathbb{X}_{m,n}$ 
is realized by an $n$-tuple of points on the elliptic curve 
$C_{\lambda,\mu}\subset\mathbb{P}^{m-1}(\mathbb{C})$ for any choice of 
$\lambda, \mu=(\mu_1,\ldots,\mu_m)$ such that 
$\vep_0=\lambda+\mu_1+\cdots+\mu_m$.  
The meromorphic mapping $\varphi_{m,n}$ is equivariant by construction 
under the action of the symmetric group 
$\mathfrak{S}_n=\br{s_1,\ldots,s_{n-1}}$. 
Also, the equivariance with respect to $s_0$ is clearly seen 
by the explicit formula \eqref{eq:uinep} 
for the $u$ coordinates. 
Hence we obtain 

\begin{theorem}\label{thm:lin}
The meromorphic mapping 
$\varphi_{m,n}: \mathfrak{h}_{m,n}\ratto 
\mathbb{X}_{m,n}$
$($resp. $\overline{\varphi}_{m,n}: E_{m,n}\ratto$ 
$\mathbb{X}_{m,n})$ defined as above is $W_{m,n}$-equivariant 
with respect to the canonical linear action of the Weyl group 
$W_{m,n}$ on 
$\mathfrak{h}_{m,n}$ and its nonlinear birational action 
on $\mathbb{X}_{m,n}$ by Cremona transformations. 
\end{theorem}
This theorem means that the $\Omega$-periodic 
functions \eqref{eq:uinep} satisfy the equations 
\begin{equation}
u_{i,j}(w(\vep))=S^{w}_{i,j}(u(\vep))\qquad(i=1,\ldots,m-1; j=m+2,\ldots,n) 
\end{equation}
for any $w\in W_{m,n}$, 
where $w(\vep)=(w(\vep_0),w(\vep_1),\ldots,w(\vep_n))$. 
Namely, \eqref{eq:uinep} give a solution of the system 
of functional equations \eqref{eq:Wmn} for the 
$u$ variables. 

As we discussed in the previous section, 
in the relative situation $\mathbb{X}_{m,n+1}\to \mathbb{X}_{m,n}$, 
the birational action of $W_{m,n}$ on $\mathbb{X}_{m,n+1}$ 
is expressed as 
\begin{equation}
w(u_{i,j})=S^w_{i,j}(u),\qquad 
w(z_i)=R^w_{i}(u;z). 
\end{equation}
By substituting $u_{ij}=u_{ij}(\vep)$ as in 
\eqref{eq:uinep}, we obtain a realization of 
$W_{m,n}$ as a group of automorphisms of the 
field $\mathcal{M}(E_{m,n})(z)$ 
of rational functions in $z$ with coefficients in the 
field of meromorphic functions on $E_{m,n}$:
\begin{equation}\label{eq:Crmn}
w(z_i)=R^w_i(\vep;z)\qquad(i=1,\ldots,m-1),
\end{equation}
where we have used the notation $R^{w}_i(\vep;z)$ again 
instead of $R^w_i(u(\vep);z)$. 
We will refer to  
this system \eqref{eq:Crmn} of functional 
equations for the $z$ variables as 
the {\em elliptic Cremona system} of type $(m,n)$. 
The action of the simple reflection $s_k$ ($k=0,1,\ldots,n-1$) 
on the $z$ variables 
is now given as follows: 
\begin{equation}\label{eq:Crmnins}
\arraycolsep=2pt
\begin{array}{lrll}
k=0: 
& s_0(z_i) &=\dfrac{1}{z_i},
\\
k=1,\ldots,m-2: 
& s_k(z_i) &=z_{s_k(i)},
\\[6pt]
k=m-1: 
& 
s_{m-1}(z_{i}) & 
=\left\{
\begin{array}{ll}
\dfrac{z_{i}}{z_{m-1}}\quad & (i=1,\ldots,m-2),\\[10pt]
\dfrac{1}{z_{m-1}} &(i=m),
\end{array}
\right.
\\[20pt]
k=m: 
& s_{m}(z_{i})&=1-z_{i},
\\[4pt]
k=m+1:
& s_{m+1}(z_{i}) &=\dfrac{z_{i}}{u_{i,m+2}(\vep)}
\\
k=m+2,\ldots,n-1: 
& s_k(z_{i}) &=z_{i}.
\end{array}
\end{equation} 
Note that the dependence on the $\vep$ variables 
enters in the action of $s_{m+1}$.
Also, from Theorem \ref{thm:lin} for the case of 
$\mathbb{X}_{m,n+1}$ with $\vep_{n+1}=t$, we see that 
the functions
\begin{equation}\label{eq:cansol}
z_i(\vep;t)= 
\dfrac{[\alpha_0+\vep_{m,m+1}][\vep_{i,m+1}]}
{[\vep_{m,m+1}][\alpha_0+\vep_{i,m+1}]}
\dfrac{[\alpha_0+\vep_{i}-t][\vep_{m}-t]}
{[\vep_{i}-t][\alpha_0+\vep_{m}-t]}
\qquad(i=1,\ldots,m-1)
\end{equation}
satisfy the functional equations 
\begin{equation}
z_i(w(\vep);t)=R^w_i(\vep;z(\vep;t))\qquad(i=1,\ldots,m-1),
\end{equation}
for any $w\in W_{m,n}$. 
Namely, \eqref{eq:cansol} provides a one-parameter 
family of solutions to the elliptic Cremona system 
\eqref{eq:Crmn} of type $(m,n)$.  
This solution will be called the {\em canonical solution}
of the elliptic Cremona system, 
which corresponds to the {\em vertical solutions} in the 
context of differential Painlev\'e equations. 

\section{Tau functions for the elliptic Cremona system}
\label{taufunctions} 
In the case of the elliptic Cremona systems as we introduced 
above, the homogeneous polynomials $F^{w}_i(x)$ and $G^{w}_i(x)$ 
in \eqref{eq:ECrmn} can be determined without ambiguity
by means of the action of the Weyl group on the {\em $\tau$-functions}. 
In this section, we introduce a framework of $\tau$-functions 
for the elliptic Cremona systems, and show that the Weyl group 
action is translated into bilinear equations of Hirota-Miwa 
type for $\tau$-functions on the lattice. 
\par\medskip
By using the natural linear action of $W_{m,n}$ on $\mathfrak{h}_{m,n}^\ast$, 
we define the set of {\em real roots} by 
$\Delta_{m,n}^{\re}=W_{m,n}\{\alpha_0,\alpha_1,\ldots,\alpha_{n-1}\}$,
and denote by $\mathbb{K}=\mathbb{C}([\alpha];\alpha\in\Delta_{m,n}^{\re})$ 
the field of meromorphic functions on $\mathfrak{h}_{m,n}$ 
generated by $[\alpha]$ for all real roots $\alpha\in\Delta_{m,n}^{\re}$. 

In order to formulate $\tau$-functions for the elliptic 
Cremona system of type $(m,n)$, 
we use two kinds of variables (indeterminates)
$f_1,\ldots,f_m$, which will be identified with a system 
of homogeneous coordinates for $\mathbb{P}^{m-1}(\mathbb{C})$, 
and $\tau_1,\ldots,\tau_n$ corresponding to the $n$ points 
$p_1,\ldots,p_n$ in the configuration $[p_1,\ldots,p_n]$. 
We denote by 
\begin{equation}
\mathcal{L}=\mathbb{K}(f_1,\ldots,f_m;\tau_1,\ldots,\tau_n)
\end{equation}
the field of rational functions in the $f$ variables and 
the $\tau$ variables with coefficients in $\mathbb{K}$. 
On this field $\mathcal{L}$, we define the automorphisms  
$s_0, s_1,\ldots s_{n-1}$ as follows. 
These elements act on the coefficient field $\mathbb{K}$ through 
the canonical 
$W_{m,n}$-action on the real roots: For each $k=0,1,\ldots,n-1$, 
\begin{equation} 
s_k([\alpha])=[s_k(\alpha)]=[\alpha-\br{h_k,\alpha}\alpha_k]\qquad
(\alpha\in\Delta^{\re}_{m,n}).
\end{equation}
The action of $s_k$ on $\tau_j$ ($j=1,\ldots,n$) is defined by
\begin{equation}\label{eq:sontau}
\begin{array}{ll}
s_0(\tau_j)=
\left\{\begin{array}{ll}
\tau_j\,f_j\quad&(j=1,\ldots,m),\\
\tau_j\quad&(j=m+1,\ldots,n),
\end{array}\right.
\\
s_k(\tau_j)=\tau_{s_k(j)}\qquad\quad
(k=1,\ldots,n-1;j=1,\ldots,n).
\end{array}
\end{equation}
The action of $s_k$ on $f_i$ ($i=1,\ldots,m$) is defined by
\begin{equation}
\arraycolsep=2pt
\begin{array}{rlll}
s_0(f_i)&=\dfrac{1}{f_i}\\[8pt]
s_k(f_i)&=f_{s_k(i)}\quad\quad&(k=1,\ldots,m-1),\\[8pt]
s_k(f_i)&=f_i\qquad\quad&(k=m+1,\ldots,n), 
\end{array}
\end{equation}
and
\begin{equation}\label{eq:smonf}
\arraycolsep=2pt
\begin{array}{rl}
s_m(f_i)&=
\dfrac{\tau_m}{\tau_{m+1}}
\dfrac{[\alpha_0+\vep_{m,m+1}][\vep_{i,m+1}]f_i-
[\alpha_0+\vep_{i,m+1}][\vep_{m,m+1}]f_m
}{[\alpha_0][\vep_{i,m}]}\\[10pt]
&\hfill (i=1,\ldots,m-1),\\[4pt]
s_m(f_m)&=
\dfrac{\tau_m}{\tau_{m+1}}f_m. 
\end{array}
\end{equation}
\begin{theorem}\label{thm:ftau}
The automorphisms $s_0,s_1,\ldots,s_{n-1}$ of 
$\mathcal{L}$ 
%$=\mathbb{K}[f_1,\ldots,f_m][\tau_1^{\pm1},\ldots,\tau_n^{\pm1}]$ 
defined as above
satisfy the fundamental relations for the simple reflections 
of the Weyl group $W_{m,n}$. 
\end{theorem}
This theorem can be proved directly by using the Riemann relation 
for $[x]$.

We remark that the action of the Weyl group $W_{m,n}$ preserves 
the subalgebra
\begin{equation}
\mathcal{R}=S[\tau_1^{\pm1},\ldots,\tau_n^{\pm1}]\subset\mathcal{L},
\end{equation}
where 
\begin{equation}
S=\bigoplus_{d\in\mathbb{Z}}\,S_d,\qquad
S_d=f_m^d\,\mathbb{K}(f_1/f_m,\ldots,f_{m-1}/f_m). 
\end{equation} 
Consider the following elements in $S_0$: 
\begin{equation}
z_i=\dfrac{a_i}{a_m}\dfrac{f_i}{f_m},\qquad
a_i=\dfrac{[\vep_{i,m+1}]}{[\alpha_0+\vep_{i,m+1}]}
\quad(i=1,\ldots,m), 
\end{equation}
so that $S_0=\mathbb{K}(z_1,\ldots,z_{m-1})$ and that 
$(z_1:\ldots:z_{m-1}:1)=(a_1f_1:\ldots:a_mf_m)$. 
Then the action of $s_k$ ($k=0,1,\ldots,n-1$) 
on these $z$ variables 
coincides with the one given earlier in \eqref{eq:Crmnins}. 
In this sense the $f$ variables are thought of as 
{\em normalized} homogeneous coordinates. 
We also remark that the $f$ variables and the $\tau$ variables 
for the canonical solution are given by 
\begin{equation}
f_i^{C}=\dfrac{[\alpha_0+\vep_i-t]}{[\vep_i-t]}
\quad(i=1,\ldots,m),\qquad
\tau_j^{C}=[\vep_j-t]
\quad(j=1,\ldots,n). 
\end{equation}
(We have used the superscript $C$ to indicate that they are 
``canonical'', and also associated with the elliptic curve $C_{\lambda,\mu}$.)

\par\medskip
It is also convenient to introduce another $\tau$ variable 
$\tau_0$ and define $x_i$ ($i=1,\ldots,m$) by the formula
\begin{equation}\label{eq:ftox}
f_i=\dfrac{\tau_0\,x_i}{\tau_1\cdots\tau_m}\qquad(i=1,\ldots,m). 
\end{equation}
To be more precise, consider the field of rational functions 
\begin{equation}
\widetilde{\mathcal{L}}=\mathbb{K}(x_1,\ldots,x_m;\tau_0,\tau_1,\ldots,\tau_m)
\end{equation}
and identify $\mathcal{L}$ with its subfield
\begin{equation}
\mathcal{L}=\mathbb{K}(\tau_0x_1,\ldots,\tau_0x_m;\tau_1,\ldots,\tau_m)
\end{equation}
by the formula \eqref{eq:ftox}. 
The action of the symmetric group 
$\mathfrak{S}_n=\br{s_1,\ldots,s_{n-1}}$ can be 
extended to $\widetilde{\mathcal{L}}$ by setting 
\begin{equation}\label{eq:skx}
s_k(x_i)=x_{s_k(i)}\quad(k=1,\ldots,m-1),
\quad
s_k(x_i)=x_i\quad(k=m+1,\ldots,n-1)
\end{equation}
for $i=1,\ldots,m$, and
\begin{equation}\label{eq:smx}
s_m(x_i)=
\dfrac{[\alpha_0+\vep_{m,m+1}][\vep_{i,m+1}]x_i-
[\alpha_0+\vep_{i,m+1}][\vep_{m,m+1}]x_m
}{[\alpha_0][\vep_{i,m}]}
\end{equation}
for $i=1,\ldots,m-1$ and $s_m(x_m)=x_m.$ 
The action of $\mathfrak{S}_n$ on the $\tau$ variables are defined as 
\begin{equation}
s_k(\tau_0)=\tau_0,\qquad s_k(\tau_j)=\tau_{s_k(j)}\quad(j=1,\ldots,n)
\end{equation}
for any $k=1,\ldots,n-1$. 
Note that the action of $s_0\in W_{m,n}$ is defined only on 
the subfield $\mathcal{L}\subset\widetilde{\mathcal{L}}$. 
The products $\tau_0\,x_i\in\mathcal{L}$ are transformed by $s_0$ as follows:
\begin{equation}
s_0(\tau_0\,x_i)=\dfrac{\tau_0^{m-1}x_1\cdots\widehat{x_i}\cdots x_m}
{(\tau_1\cdots\tau_m)^{m-2}}\qquad(i=1,\ldots,m). 
\end{equation}

We regard $x=(x_1,\ldots,x_m)$ as a normalized homogeneous 
coordinate system such that 
%$z_i=a_ix_i/a_mx_m$ ($i=1,\ldots,m-1$), 
$(z_1:\ldots:z_{m-1}:1)=(a_1x_1:\ldots:a_mx_m)$. 
%\begin{equation}
%z_i=\dfrac{[\alpha_0+\vep_{m,m+1}]
%[\vep_{i,m+1}]}{[\vep_{m,m+1}][\alpha_0+\vep_{i,m}]}
%\dfrac{x_i}{x_m}\qquad(i=1,\ldots,m-1),
%\end{equation}
Note that the canonical solution is given by 
\begin{equation}
\tau_0\,x_i=
x_i^C(t)=
[\alpha_0+\vep_i-t]\dprod{1\le k\le m;k\ne i}{}[\vep_k-t]
\qquad(i=1,\ldots,m). 
\end{equation}
Accordingly, we define the $x$ coordinates of the reference points 
$p_1,\ldots,p_n$ by 
\begin{equation}\label{eq:xpj}
x(p_j)=(x_1^{C}(\vep_j):\ldots:x_m^C(\vep_j)), 
\qquad 
x_i^C(\vep_j)=
[\alpha_0+\vep_{i,j}]\dprod{1\le k\le m;k\ne i}{}[\vep_{k,j}],
\end{equation}
for $j=1,\ldots,n$. 
For each element 
\begin{equation}\label{eq:Lambda}
\Lambda=de_0-\nu_1e_1-\cdots-\nu_n e_n\in L_{m,n}
\qquad(d,\nu_1,\ldots,\nu_n\in\mathbb{Z}),
\end{equation}
we define the {\em formal exponential} $\tau^\Lambda$ by
\begin{equation}
\tau^\Lambda=\tau_0^d\,\tau_1^{-\nu_1}\,\cdots\,\tau_{n}^{-\nu_n}. 
\end{equation}
Since $f_i=\tau^{h_0}x_i$ ($i=1,\ldots,m$), 
the algebra $\mathcal{R}$ can be expressed alternatively as 
\begin{equation}
\mathcal{R}=
\bigoplus_{\Lambda\in L_{m,n}} 
\mathbb{K}(x)_{\deg(\Lambda)}\,\tau^\Lambda, 
\qquad \mathbb{K}(x)_d=x_m^d\,\mathbb{K}(x_1/x_m,\ldots,x_{m-1}/x_m), 
\end{equation}
where $\deg(\Lambda)$ stands for the coefficient $d$ of 
$e_0$ in \eqref{eq:Lambda}. 
In the $x$ and $\tau$ variables, the action of the Weyl group 
on $\mathcal{R}$ is described as follows: 
\begin{equation}
s_0(\tau^\Lambda\varphi(x))=\tau^{s_0.\Lambda}\,
x_1^{d-\nu_1}\cdots x_m^{d-\nu_m}\,{}^{s_0}\!\varphi(x^{-1})
\end{equation}
and 
\begin{equation}
s_k(\tau^{\Lambda}\varphi(x))=\tau^{s_k.\Lambda}\,{}^{s_k}\!\varphi(s_k(x))
\qquad(k=1,\ldots,n-1)
\end{equation}
for any $\varphi(x)=\varphi(x_1,\ldots,x_m)\in\mathbb{K}(x)_d$, where 
${}^w\!\varphi(x)$ stands for the polynomial obtained from $\varphi(x)$ 
by applying $w$ to its coefficients. 
The expression $s_k(x)=(s_k(x_1),\ldots,s_k(x_m))$ 
($k=1,\ldots,n-1$) should be understood in the sense of the action of 
$\mathfrak{S}_n$ on $\widetilde{\mathcal{L}}$ described in
\eqref{eq:skx} and \eqref{eq:smx}.  

\par\medskip
We are now ready to introduce the lattice $\tau$-functions for the 
elliptic Cremona system of type $(m,n)$.
We consider the $W_{m,n}$-orbit of the lattice point $e_n$ 
in $L_{m,n}$:
\begin{equation}
M_{m,n}=W_{m,n}e_n=W_{m,n}\{e_1,\ldots,e_n\}\subset L_{m,n}. 
\end{equation}
Notice that the stabilizer of $e_n$ is $W_{m,n-1}$, 
and that $\tau_n$ is $W_{m,n-1}$-invariant. 
Hence, for each $\Lambda\in M_{m,n}$ 
we can define an element $\tau(\Lambda)=w(\tau_n)\in\mathcal{R}$
by taking any element $w\in W_{m,n}$ such that $w.e_n=\Lambda$. 
This family of $(\tau(\Lambda))_{\Lambda\in M_{m,n}}$ 
of elements of $\mathcal{R}$, indexed by the lattice points 
$M_{m,n}\subset L_{m,n}$ will be called the {\em lattice 
$\tau$-functions} for the elliptic Cremona system of type $(m,n)$.
These $\tau$-functions are determined by the 
condition $\tau(e_j)=\tau_j$ $(j=1,\ldots,n)$
together with the compatibility condition 
\begin{equation}\label{eq:WM}
w(\tau(\Lambda))=\tau(w.\Lambda)\qquad
(\Lambda\in M_{m,n}; w\in W_{m,n})
\end{equation}
with respect to the action of $W_{m,n}$ on $M_{m,n}$. 
By using 
\begin{equation}
f_i=\dfrac{s_0(\tau_i)}{\tau_i}=\dfrac{\tau(e_i+h_0)}{\tau(e_i)}
\qquad(i=1,\ldots,m), 
\end{equation}
from the action of $s_m$ \eqref{eq:smonf}
on the $f$ variables we obtain the following bilinear 
relations for the lattice $\tau$-functions:
\begin{equation}
\begin{array}{ll}
[\alpha_0][\vep_{i,m}]\tau(h_0+e_i+e_m-e_{m+1})\tau(e_{m+1})\\
=
[\alpha_0+\vep_{m,m+1}][\vep_{i,m+1}]\tau(h_0+e_i)\tau(e_{m})\\
\quad\mbox{}-[\alpha_0+\vep_{i,m+1}][\vep_{m,m+1}]\tau(h_0+e_m)\tau(e_{i}). 
\end{array}
\end{equation}
From this we obtain
\begin{theorem}\label{thm:HirotaMiwa}
The lattice $\tau$-functions $(\tau(\Lambda))_{\Lambda\in M_{m,n}}$ 
defined as above satisfy the following bilinear equations of 
Hirota-Miwa type\,$:$  For any choice of mutually distinct indices 
$i,j,k, l_1,\ldots,l_{m-2}\in\{1,\ldots,n\}$, 
\begin{equation}\label{eq:HM}
\begin{array}{c}
[\vep_{j,k}][\lambda-\vep_j-\vep_k]\tau(e_i)\tau(\Lambda-e_i)
+ 
[\vep_{k,i}][\lambda-\vep_k-\vep_i]\tau(e_j)\tau(\Lambda-e_j)\\[4pt]
\mbox{}+[\vep_{i,j}][\lambda-\vep_i-\vep_j]\tau(e_k)\tau(\Lambda-e_k)=0,
\end{array}
\end{equation}
where $\Lambda=e_0-e_{l_1}-\cdots-e_{l_{m-2}}$ and 
$\lambda=\ipr{\Lambda}{\cdot}=\vep_0-\vep_{l_1}-\cdots-\vep_{l_{m-2}}$. 
\end{theorem}
\begin{figure}
$$
%\fbox{
\begin{picture}(360,205)(-40,-40)
\unitlength=0.5pt
\thicklines
\multiput(0,0)(4,0){50}{\line(1,0){2}}
\put(0,0){\line(5,-3){100}}
\put(200,0){\line(-5,-3){100}}
\put(0,0){\line(1,1){200}}
\put(200,0){\line(-1,1){200}}
\put(0,200){\line(1,0){200}}
\put(0,200){\line(5,3){100}}
\put(200,200){\line(-5,3){100}}
\multiput(100,100)(0,4){40}{\line(0,1){2}}
\put(100,100){\line(0,-1){160}}
\put(-40,200){$\tau(a)$}
\put(205,200){$\tau(b)$}
\put(90,270){$\tau(c)$}
\put(-75,0){$\tau(\Lambda-b)$}
\put(205,0){$\tau(\Lambda-a)$}
\put(70,-80){$\tau(\Lambda-c)$}
\put(250,80){$\begin{array}{ll}
\quad\,[\beta-\gamma][\lambda-\beta-\gamma]\ \tau(a)\tau(\Lambda-a)\\[4pt]
\quad+\ [\gamma-\alpha][\lambda-\gamma-\alpha]\ \tau(b)\tau(\Lambda-b)\\[4pt]
\quad+\ [\alpha-\beta][\lambda-\alpha-\beta]\ \tau(c)\tau(\Lambda-c)=0
\end{array}$}
\end{picture}
%}
$$
\caption{Bilinear equations for the lattice $\tau$-functions}
\end{figure}
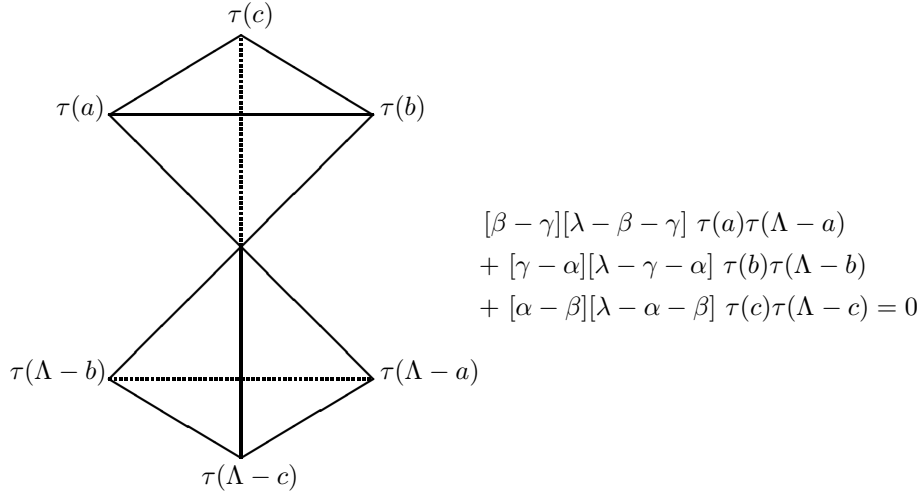
The lattice $\tau$-functions for 
the canonical solution are given simply by
\begin{equation}\label{eq:taunorm}
\tau^C(\Lambda)=[\lambda-t]\qquad \mbox{with}\quad
\lambda=\ipr{\Lambda}{\cdot}\in\mathfrak{h}_{m,n}^{\ast}
\end{equation}
for any $\Lambda\in M_{m,n}$. In this case the bilinear equations 
in Theorem \ref{thm:HirotaMiwa} recover the Riemann relation for 
the function $[x]$. 

We also remark that these bilinear equations of Hirota-Miwa type 
characterize the lattice $\tau$-functions for the elliptic 
Cremona system. To be more precise, 
suppose that the natural action of $W_{m,n}$ on $\mathbb{K}$ 
is extended to a field $\mathcal{K}$ containing $\mathbb{K}$
(as that of a group of automorphisms). 
If a family of nonzero elements $(\tau(\Lambda))_{\Lambda\in M_{m,n}}$ 
of $\mathcal{K}$, indexed by $M_{m,n}$, satisfies the two conditions 
\eqref{eq:WM} and \eqref{eq:HM}, then the elements 
\begin{equation}
f_i=\dfrac{\tau(e_i+h_0)}{\tau(e_i)}\quad(i=1,\ldots,m),\qquad
\tau_j=\tau(e_j)\quad(j=1,\ldots,n)
\end{equation}
of $\mathcal{K}$ recover the relations \eqref{eq:sontau}--\eqref{eq:smonf} 
for the action of $W_{m,n}$ on $\mathcal{L}$. 

\par\medskip
Recall that the algebra $\mathcal{R}$ is expressed as 
\begin{equation}
\mathcal{R}=
\bigoplus_{\Lambda\in L_{m,n}} 
\mathbb{K}(x)_{\deg(\Lambda)}\,\tau^\Lambda. 
\end{equation}
As in Section \ref{Cremona}, for each $\Lambda\in L_{m,n}$ 
of the form \eqref{eq:Lambda} 
we define the 
$\mathbb{K}$-vector subspace $L(\Lambda)\subset\mathbb{K}[x]_d$ 
by 
\begin{equation}
L(\Lambda)=\{f(x)\in \mathbb{K}[x]_d\ |\  \mbox{ord}_{p_j}f(x)\ge\nu_j 
\quad(j=1,\ldots,n)\},
\end{equation}
with the reference points $p_j$ specified by the $x$ coordinates 
as in \eqref{eq:xpj}. 
Then one can show that the subalgebra
\begin{equation}
\mathcal{S}=
\bigoplus_{\Lambda\in L_{m,n}} 
L(\Lambda)\,\tau^\Lambda\subset \mathcal{R} 
\end{equation}
is preserved under the action of $W_{m,n}$.  In fact 
each $w\in W_{m,n}$ induces a $\mathbb{C}$-isomorphism 
\begin{equation}
w.:  L(\Lambda)\,\tau^{\Lambda} \iso L(w.\Lambda)\,\tau^{w.\Lambda}
\end{equation}
for any $\Lambda\in L_{m,n}$. 
(This fact can be established by examining the 
cases of simple reflections $s_0,s_1,\ldots,s_{n-1}$.)
Since $\tau_j\in L(e_j)\tau^{e_j}$ ($j=1,\ldots,n$), 
the lattice $\tau$-function $\tau(\Lambda)$ for $\Lambda\in M_{m,n}$
can be expressed in the form
\begin{equation}\label{eq:tauphi}
\tau(\Lambda)
=\tau^{\Lambda}\,\phi(\Lambda;x),
\qquad\phi(\Lambda;x)\in L(\Lambda)
\end{equation}
in terms of the original $\tau$ variables $\tau_0,\tau_1,\ldots,
\tau_n$ and the homogeneous coordinates $x_1,\ldots,$ $x_m$. 
We remark that this family of homogeneous polynomials 
$\phi(\Lambda;x)$ ($\Lambda\in M_{m,n}$) is determined uniquely 
by the following properties:
\begin{equation}\label{eq:phirec}
\begin{array}{ll}
\phi(e_j;x)=1\qquad&(j=1,\ldots,n),\\[4pt]
\phi(s_0.\Lambda;x)=x_1^{d-\nu_1}\cdots x_m^{d-\nu_m}\,
{}^{s_0}\!\phi(\Lambda;x^{-1}),\\[4pt]
\phi(s_k.\Lambda;x)={}^{s_k}\phi(\Lambda;s_k(x))
\qquad\qquad&(k=1,\ldots,n-1),
\end{array} 
\end{equation}
where $\Lambda=de_0-\nu_1e_1-\ldots-\nu_ne_n$. 
We sometimes refer to this family of polynomials $\phi(\Lambda;x)$  
as the {\em $\tau$-cocycle}. 
They correspond to what are called {\em $\phi$-factors} in \cite{Noumi}. 

Since $f_i=\tau^{h_0}x_i$ $(i=1,\ldots,m)$, the 
formula \eqref{eq:tauphi} is rewritten also into
\begin{equation}
\tau(\Lambda)
=\tau^{\Lambda-\deg(\Lambda)e_0}\,\phi(\Lambda;\tau_0\,x)
=\tau^{\Lambda-\deg(\Lambda)h_0}\,\phi(\Lambda;f)
\end{equation}
in terms of the variables $\tau_0x_i$ or $f_i$. 
For $\Lambda$ as in \eqref{eq:Lambda}, we have 
\begin{equation} 
\tau(\Lambda)=
\dfrac{\phi(\Lambda;\tau_0x)}{\tau_1^{\nu_1}\cdots\tau_n^{\nu_n}} 
=
\dfrac{\tau_1^{d-\nu_1}\cdots\tau_m^{d-\nu_m}}
{\tau_{m+1}^{\nu_{m+1}}\cdots\tau_n^{\nu_n}}\, 
\phi(\Lambda;f). 
\end{equation}
Note here that $\phi(\Lambda;x)$ is a homogeneous polynomial of 
degree $d$ having a zero of multiplicity $\ge \nu_j$ 
at each $p_j$ ($j=1,\ldots,n$). 
In the case of the canonical solution, 
the above formula gives rise to  
\begin{equation}
\phi(\Lambda;x^C(t))=[\lambda-t][\vep_1-t]^{\nu_1}\cdots[\vep_n-t]^{\nu_n},
\qquad\lambda=\ipr{\Lambda}{\cdot}. 
\end{equation}
In this form, it is clearly seen that $\phi(\Lambda;x^{C}(t))$ 
has zeros at $p_1,\ldots,p_n$ with expected multiplicities. 

Note that 
\begin{equation}
f_i=
\dfrac{\tau(e_i+h_0)}{\tau(e_i)}
=\dfrac{\tau(s_0.e_i)}{\tau(e_i)}\qquad(i=1,\ldots,m)
\end{equation}
implies 
\begin{equation}
w(f_i)
=\dfrac{\tau(ws_0.e_i)}{\tau(w.e_i)}
=\dfrac{\tau^{ws_0.e_i}}{\tau^{w.e_i}}
\dfrac{\phi(ws_0.e_i;x)}{\phi(w.e_i;x)}
=\tau^{w.h_0} \dfrac{\phi(ws_0.e_i;x)}{\phi(w.e_i;x)}
\end{equation}
for any $w\in W_{m,n}$.
Hence, the action of $w\in W_{m,n}$ on the variables 
$f_1,\ldots,f_m$ and 
$\tau_1,\ldots,\tau_n$ 
can be written in a closed form as follows:
\begin{equation}
\arraycolsep=2pt
\begin{array}{rll}
w(f_i)&=\tau^{w.h_0-\deg(w.h_0) h_0}\,
\dfrac{\phi(ws_0.e_i;f)}{\phi(w.e_i;f)}\qquad&(i=1,\ldots,m),\\[8pt]
w(\tau_j)&=\tau^{w.e_j-\deg(w.e_j) h_0}\,
\phi(w.e_j;f)\qquad&(j=1,\ldots,n). 
\end{array}
\end{equation}
Also, from 
\begin{equation}
(z_1:\ldots:z_{m-1}:1)=(a_1f_1:\ldots:a_mf_m), \qquad
a_i=\dfrac{[\vep_{i,m+1}]}{[\alpha_0+\vep_{i,m+1}]}, 
\end{equation}
we obtain 
\begin{equation}
(w(z_1):\ldots:w(z_{m-1}):1)
=
\left(
w(a_1)
\dfrac{\phi(ws_0.e_1;x)}{\phi(w.e_1;x)}:
\ldots:
w(a_m)
\dfrac{\phi(ws_0.e_m;x)}{\phi(w.e_m;x)}\right)
\end{equation}
for any $w\in W_{m,n}$. 
This formula provides a refinement of Theorem \ref{thm:GF} 
for the elliptic Cremona system of type $(m,n)$. 
\comment{
Namely one has
\begin{equation}
w(z_i)=R^w_i(z)=
\dfrac{c_i}{c_m}
\dfrac{G^w_i(z)F^w_m(z)}{F^w_i(z)G^w_m(z)}\qquad(i=1,\ldots,m-1),
\end{equation}
where 
$c_i=[w(\vep_{i,m+1})]/[ws_0(\vep_{i,m+1})]$
and
\begin{equation}
F^w_i(z)=(x_m)^{-\deg(w(e_i))}\phi(w.e_i;x),\quad
G^w_i(z)=(x_m)^{-\deg(ws_0(e_i))}\phi(ws_0.e_i;x)
\end{equation}
for $i=1,\ldots,m$. 
}

\section{Elliptic difference Painlev\'e equation}
\label{ellipticdifference}

In the rest of this article, we confine ourselves to the 
elliptic Cremona system of type $(3,9)$ which has 
the affine Weyl group symmetry of type $E^{(1)}_8$. 
The discrete dynamical system arising from the  
lattice part of $W(E^{(1)}_8)$ is the elliptic difference 
Painlev\'e equation.
\par\medskip
The parameter space for the elliptic Cremona system of type $(3,9)$ 
is the $10$-dimensional vector space 
\begin{equation}
\mathfrak{h}_{3,9}=
\mathbb{C}e_0\oplus
\mathbb{C}e_1\oplus\cdots\oplus
\mathbb{C}e_9
\end{equation}
with the nondegenerate symmetric bilinear form $\ipr{\cdot}{\cdot}: 
\mathfrak{h}_{3,9}\times\mathfrak{h}_{3,9}\to\mathbb{C}$ 
such that 
\begin{equation}
\begin{array}{ll}
\ipr{e_0}{e_0}=-1,\qquad &\ipr{e_j}{e_j}=1\quad(j=1,\ldots,9),\\
\ipr{e_i}{e_j}=0\qquad & (i,j\in\{1,\ldots,9\}; i\ne j), 
\end{array}
\end{equation} 
which we regard as the complexification of the lattice 
$L_{3,9}=\mathbb{Z}e_0\oplus
\mathbb{Z}e_1\oplus\cdots\oplus
\mathbb{Z}e_9.$ 
We take the linear functions $\vep_j=\ipr{e_j}{\cdot}$ 
($j=0,1,\ldots,9$) as the canonical coordinates for $\mathfrak{h}_{3,9}$,
so that $\mathfrak{h}_{3,9}^\ast=
\mathbb{C}\vep_0\oplus
\mathbb{C}\vep_1\oplus\cdots\oplus
\mathbb{C}\vep_9$. 
The {\em simple coroots} $h_j\in\mathfrak{h}_{3,9}$ and 
the {\em simple roots} $\alpha_j=\ipr{h_j}{\cdot}\in\mathfrak{h}_{3,9}^\ast$ 
($j=0,1,\ldots,8$) for this case are 
\begin{equation}
\begin{array}{ll}
h_0=e_0-e_1-e_2-e_3,\qquad h_j=e_j-e_{j+1}\quad(j=1,\ldots,8),\\
\alpha_0=\vep_0-\vep_1-\vep_2-\vep_3,\qquad 
\alpha_j=\vep_j-\vep_{j+1}\quad(j=1,\ldots,8). 
\end{array}
\end{equation}
The $9\times 9$ matrix $(\br{h_i,\alpha_j})_{i,j=0}^{8}$ is the 
Cartan matrix of type $E^{(1)}_8$ with the following Dynkin diagram. 
\begin{equation}
%\fbox{
%\unitlength=1.2pt
\begin{picture}(205,20)(-35,0)
\put(20,0){\circle{4}}
\put(22,0){\line(1,0){16}}
\put(40,0){\circle{4}}
\put(42,0){\line(1,0){16}}
\put(60,0){\circle{4}}
\put(62,0){\line(1,0){16}}
\put(80,0){\circle{4}}
\put(82,0){\line(1,0){16}}
\put(100,0){\circle{4}}
\put(102,0){\line(1,0){16}}
\put(120,0){\circle{4}}
\put(122,0){\line(1,0){16}}
\put(140,0){\circle{4}}
\put(142,0){\line(1,0){16}}
\put(160,0){\circle{4}}
%\put(162,0){\line(1,0){16}}
%\put(180,0){\circle{4}}
\put(60,2){\line(0,1){14}}
\put(60,18){\circle{4}}
\put(18,-8){\small$\alpha_{1}$}
\put(38,-8){\small$\alpha_{2}$}
\put(58,-8){\small$\alpha_{3}$}
\put(78,-8){\small$\alpha_{4}$}
\put(98,-8){\small$\alpha_{5}$}
\put(118,-8){\small$\alpha_{6}$}
\put(138,-8){\small$\alpha_{7}$}
\put(158,-8){\small$\alpha_{8}$}
\put(64,18){\small$\alpha_0$}
\put(-35,-2){$E^{(1)}_8 :$}
\end{picture}
%}
\vspace{8pt}
\end{equation}
We denote by 
\begin{equation}
Q_{3,9}=\mathbb{Z}\alpha_0\oplus
\mathbb{Z}\alpha_1\oplus\cdots\oplus 
\mathbb{Z}\alpha_8\subset \mathfrak{h}_{3,9}^\ast,
\quad 
Q_{3,9}\iso Q(E^{(1)}_8), 
\end{equation}
the corresponding root lattice. 
The affine Weyl group 
$W_{3,9}=\br{s_0,s_1,\ldots,s_8}$ of type $E^{(1)}_8$ 
acts in a standard way on 
$\mathfrak{h}_{3,9}$ and $\mathfrak{h}_{3,9}^\ast$, 
so that 
the symmetric bilinear forms of $\mathfrak{h}_{3,9}$ and 
$\mathfrak{h}_{3,9}^\ast$ are both $W_{3,9}$-invariant. 
The {\em canonical central element} 
\begin{equation} 
c=3e_0-e_1-\cdots-e_9\in\mathfrak{h}_{3,9}
\end{equation}
is orthogonal to all $h_j$ ($j=0,1,\ldots,8$), and hence 
$W_{3,9}$-invariant. 
The corresponding $W_{3,9}$-invariant element in $\mathfrak{h}_{3,9}^\ast$
\begin{equation}
\delta=\ipr{c}{\cdot}
=3\vep_0-\vep_1-\cdots-\vep_9\in\mathfrak{h}_{3,9}^\ast
\end{equation}
is called the {\em null root}; it plays the role of  
the scaling constant for difference equations in the 
context of discrete Painlev\'e equation. 
The set $\Delta_{3,9}^{\re}=W_{3,9}\{\alpha_0,\alpha_1,\ldots,\alpha_8\}$ 
of {\em real roots} is now given by 
\begin{equation}
\begin{array}{ll} 
\Delta_{3,9}^{\re}=
\{\pm \vep_{ij}+ n\delta \ |\  1\le i<j\le 9,\ n\in\mathbb{Z}\}\\
\qquad\quad
\cup\ 
\{\pm \vep_{ijk}+ n\delta \ |\  1\le i<j<k\le 9,\ n\in\mathbb{Z}\},
\end{array} 
\end{equation}
where $\vep_{ij}=\vep_i-\vep_j$ and $\vep_{ijk}=\vep_0-\vep_i-\vep_j-\vep_k$ 
for $i,j,k\in\{1,\ldots,9\}$.
For each real root $\alpha\in\Delta_{3,9}^{\re}$, 
the reflection $s_\alpha: \mathfrak{h}_{3,9}^\ast\to\mathfrak{h}_{3,9}^\ast$  
with respect to $\alpha$ is defined by
\begin{equation}
s_\alpha(\lambda)=\lambda-\ipr{\alpha}{\lambda}\alpha
\qquad(\lambda\in\mathfrak{h}_{3,9}^\ast). 
\end{equation}
Note also $ws_\alpha w^{-1}=s_{w.\alpha}$ for any 
$\alpha\in\Delta_{3,9}^{\re}$ and $w\in W_{3,9}$. 
When $\alpha=\vep_{ij}$ or $\vep_{ijk}$, 
we denote the reflection $s_\alpha$ simply by 
$s_{ij}$ or  $s_{ijk}$, respectively. 

\par\medskip
Following \cite{Kac}, 
for each $\alpha\in Q_{3,9}$ we define the translation 
$T_{\alpha}: \mathfrak{h}_{3,9}^\ast\to\mathfrak{h}_{3,9}^\ast$ 
by
\begin{equation}
T_{\alpha}(\lambda)=\lambda+\ipr{\delta}{\lambda}\alpha-
\left(
\frac{1}{2}\ipr{\alpha}{\alpha}\ipr{\delta}{\lambda}
+\ipr{\alpha}{\lambda}\right)\delta
\qquad(\lambda\in\mathfrak{h}_{3,9}^\ast). 
\end{equation}
Note that 
\begin{equation}
\begin{array}{ll} 
T_{\alpha}(\beta)=\beta-\ipr{\alpha}{\beta}\,\delta\quad
&(\alpha,\beta\in Q_{3,9}),\\
T_{\alpha}T_{\beta}=T_{\beta}T_{\alpha}=T_{\alpha+\beta}\qquad
&(\alpha,\beta\in Q_{3,9}),\\ 
w T_{\alpha} w^{-1}=T_{w.\alpha}\qquad 
&(\alpha\in Q_{3,9}$ and $w\in W_{3,9}). 
\end{array}
\end{equation}
For any real root $\alpha\in\Delta_{3,9}^{\re}$, 
the translation $T_{\alpha}$ can be expressed 
as the composition of two reflections 
$T_{\alpha}=s_{\delta-\alpha}s_\alpha$. 
From this fact it follows that $T_{\alpha}\in W_{3,9}$ for 
any $\alpha\in Q_{3,9}$.
Furthermore, it is known that 
the affine Weyl group $W_{3,9}$ is 
decomposed into the semidirect product of 
the root lattice 
\begin{equation}
Q_{3,8}=\mathbb{Z}\alpha_0\oplus\mathbb{Z}\alpha_1\oplus
\cdots\oplus\mathbb{Z}\alpha_7\subset Q_{3,9},
\qquad Q_{3,8}\iso Q(E_8), 
\end{equation} 
and the finite Weyl group 
$W_{3,8}=\br{s_0,s_1,\ldots,s_7}$ of type $E_8$ acting on $Q_{3,8}$. 
In fact the correspondence 
$(\alpha,w)\mapsto T_{\alpha}w$ induces the isomorphism 
$Q_{3,8}\rtimes W_{3,8}\iso W_{3,9}$ (see \cite{Kac}). 
We remark that the action of $T_{\alpha}$ on 
$\mathfrak{h}_{3,9}$ is expressed in the form 
\begin{equation}
T_{\alpha}(\Lambda)=\Lambda+\ipr{c}{\Lambda} h-
\left(
\frac{1}{2}\ipr{h}{h}\ipr{c}{\Lambda}
+\ipr{h}{\Lambda}\right)c
\qquad(\Lambda\in\mathfrak{h}_{3,9}) 
\end{equation}
by using the element $h\in\mathfrak{h}_{3,9}$ such that 
$\alpha=\ipr{h}{\cdot}$. 
When $\alpha=\vep_{ij}$ or $\vep_{ijk}$, 
we write the translation 
$T_{\alpha}$ simply as $T_{ij}$ or $T_{ijk}$, respectively. 
In this case of $(m,n)=(3,9)$, the $W_{3,9}$-orbit 
$M_{3,9}=W_{3,9}\{e_1,\ldots,e_9\}\subset L_{3,9}$
can be characterized as follows:
\begin{equation}
M_{3,9}=\{\Lambda\in L_{3,9}\ |\ \ipr{c}{\Lambda}=-1,\ \ 
\ipr{\Lambda}{\Lambda}=1\}. 
\end{equation}
Also, the correspondence $\alpha\mapsto T_{\alpha}.e_9$ 
induces a bijection $Q_{3,8}\iso M_{3,9}$. 
Any element $\Lambda\in M_{3,9}$ can be 
expressed in the form
\begin{equation}
\Lambda=de_0-\nu_1e_1-\ldots-\nu_9e_9,
\qquad 
d\ge0,\quad \nu_j\ge -1 \ \ (j=1,\ldots,9). 
\end{equation}
In fact, except for the cases $\Lambda=e_k$ ($k=1,\ldots,9$), 
the coefficients $\nu_j$ are all nonnegative. 
\par\medskip
As in the previous section, we consider the $\mathbb{K}$-algebra
\begin{equation}
\mathcal{R}=
S[\tau_1^{\pm1},\ldots,\tau_9^{\pm1}],\qquad
S=\bigoplus_{d\in\mathbb{Z}}\, f_3^d\,\mathbb{K}(f_1/f_3,f_2/f_3)
\end{equation}
of $f$ variables and $\tau$ variables. 
The standard Cremona transformation $s_0$ with respect to 
$(p_1,p_2,p_3)$ acts on the $f$ variables and $\tau$ variables 
as 
\begin{equation}
s_0(f_i)=\dfrac{1}{f_i}\quad(i=1,2,3),
\qquad 
s_0(\tau_j)=\left\{\begin{array}{ll}
\tau_j f_j \quad & (j=1,2,3),\\
\tau_j &(j=4,5,6,7,8,9).
\end{array}\right.
\end{equation}
Among the $8$ adjacent transpositions $s_k$ ($k=1,\ldots,8)$, 
$s_3$ acts nontrivially on the $f$ variables: 
\begin{equation}
\arraycolsep=2pt
\begin{array}{rll}
s_3(f_1)&=\dfrac{\tau_3}{\tau_4}
\dfrac{[\vep_{124}][\vep_{14}]f_1-[\vep_{234}][\vep_{34}]f_3}
{[\vep_{123}][\vep_{13}]},\\[10pt]
s_3(f_2)&=\dfrac{\tau_3}{\tau_4}
\dfrac{[\vep_{124}][\vep_{24}]f_1-[\vep_{134}][\vep_{34}]f_3}
{[\vep_{123}][\vep_{23}]},\\[10pt]
s_3(f_3)&=\dfrac{\tau_3}{\tau_4} f_3. 
\end{array}
\end{equation}
(Note that $\alpha_0=\vep_{123}$.)
Recall that 
the lattice $\tau$-functions $(\tau(\Lambda))_{\Lambda\in M_{3,9}}$ 
are defined as a family of dependent variables indexed by 
the $W_{3,9}$-orbit 
\begin{equation}
M_{3,9}=W_{3,9}\{e_1,\ldots,e_9\}
=\{\Lambda\in L_{3,9}\ |\ \ipr{c}{\Lambda}=-1,\ \ \ipr{\Lambda}{\Lambda}=1\}. 
\end{equation}
These $\tau$-functions $\tau(\Lambda)$ 
are characterized by the consistency condition 
\begin{equation} 
\tau(e_j)=\tau_j\quad(j=1,\ldots,9),\qquad
w(\tau(\Lambda))=\tau(w.\Lambda)\quad(\Lambda\in M_{3,9}; w\in W_{3,9})
\end{equation}
with respect to the action of $W_{3,9}$ on $M_{3,9}$, 
and the bilinear equations 
\begin{equation}
\begin{array}{c}
[\vep_{jk}][\vep_{jkl}]\tau(e_i)\tau(e_0-e_i-e_l)
+
[\vep_{ki}][\vep_{kil}]\tau(e_j)\tau(e_0-e_j-e_l)\\[4pt]
\mbox{}+
[\vep_{ij}][\vep_{ijl}]\tau(e_k)\tau(e_0-e_k-e_l)
=0
\end{array}
\end{equation}
for any mutually distinct $i,j,k,l\in\{1,\ldots,9\}$.
These bilinear equations guarantee the equivalence 
of our formulation of the elliptic difference Painlev\'e equation 
of type $E^{(1)}_8$ to that of Ohta-Ramani-Grammaticos \cite{ORG} 
on the $E_8$ lattice. 

\par\medskip
We already know that 
the lattice $\tau$-functions can be expressed in the form
\begin{equation}
\tau(\Lambda)=\tau^{\Lambda}\,\phi(\Lambda;x)=
\tau^{\Lambda-\deg(\Lambda)h_0}\phi(\Lambda;f),
\quad \phi(\Lambda;x)\in L(\Lambda). 
\end{equation}
where the $x$ coordinates are defined by $f_i=\tau^{h_0}x_i$ $(i=1,2,3)$. 
When $\Lambda=de_0-\nu_1e_1-\cdots-\nu_9e_9$, 
$\phi(\Lambda;x)$ is a homogeneous polynomial of degree 
$d=\deg(\Lambda)$, and for each $j=1,\ldots,9$ it 
has a zero of multiplicity $\ge\nu_j$ at $p_j$; 
the $x$ coordinates of $p_j$ are now given by 
\begin{equation}\label{eq:xpj3}
x(p_j)=
([\vep_{23j}][\vep_{2j}][\vep_{3j}]:
[\vep_{13j}][\vep_{1j}][\vep_{3j}]:
[\vep_{12j}][\vep_{1j}][\vep_{2j}])\qquad(j=1,\ldots,9). 
\end{equation}
By the homogeneous polynomials $\phi(\Lambda;x)$, 
the action of $W_{3,9}$ on the algebra $\mathcal{R}$
is described as
\begin{equation}\label{eq:wftau}
\arraycolsep=2pt
\begin{array}{rll} 
w(f_i)&=\tau^{w.h_0-\deg(w.h_0) h_0}\,
\dfrac{\phi(ws_0.e_i;f)}{\phi(w.e_i;f)}\qquad&(i=1,2,3),\\[8pt]
w(\tau_j)&=\tau^{w.e_j-\deg(w.e_j) h_0}\,
\phi(w.e_j;f)\qquad&(j=1,\ldots,9). 
\end{array}
\end{equation}
Recall that the $z$ variables $z=(z_1,z_2)$ are 
recovered from the $f$ variables by the formula
\begin{equation} \label{eq:zinf}
(z_1:z_2:1)=\left(
\dfrac{[\vep_{14}]}{[\vep_{234}]}f_1:
\dfrac{[\vep_{24}]}{[\vep_{134}]}f_2:
\dfrac{[\vep_{34}]}{[\vep_{124}]}f_3\right), 
\end{equation}
and hence 
\begin{equation}
(w(z_1):w(z_2):1)=\left(
\dfrac{[w(\vep_{14})]}{[w(\vep_{234})]}w(f_1):
\dfrac{[w(\vep_{24})]}{[w(\vep_{134})]}w(f_2):
\dfrac{[w(\vep_{34})]}{[w(\vep_{124})]}w(f_3)\right)
\end{equation}
for any $w\in W_{3,9}$. 
This implies 
\begin{equation} \label{eq:wzvar}
\arraycolsep=2pt 
\begin{array}{rl}
w(z_1)&=
\dfrac{[w(\vep_{14})][w(\vep_{124})]}{[w(\vep_{234})][w(\vep_{34})]}
\dfrac{G^w_1(z)F^w_3(z)}{F^w_1(z)G^w_3(z)},\\[12pt]
w(z_2)&=
\dfrac{[w(\vep_{24})][w(\vep_{124})]}{[w(\vep_{134})][w(\vep_{34})]}
\dfrac{G^w_2(z)F^w_3(z)}{F^w_2(z)G^w_3(z)}, 
\end{array}
\end{equation} 
where $F^w_i(z)$ and $G^w_i(z)$ are 
the inhomogenizations
of $\phi(w.e_i;f)$ and $\phi(ws_0.e_i;f)$, respectively,
by \eqref{eq:zinf}. 
The formulas \eqref{eq:wftau} (resp. \eqref{eq:wzvar}) 
for the translations $w=T_{\alpha}$ $(\alpha\in Q_{3,9})$ 
give the elliptic difference Painlev\'e equation 
of type $(3,9)$ in the homogeneous form of $f$ and $\tau$ 
variables (resp. in the inhomogeneous form in $z$ variables). 
In the following, we will mainly work with the homogeneous 
form \eqref{eq:wftau}. 

\par\medskip

Note that, for each $\Lambda\in M_{3,9}$, 
the homogeneous polynomial 
$\phi(\Lambda;x)$ is characterized 
by its degree and the multiplicity of zeros at 
$p_1,\ldots,p_9$.  
Thanks to this fact, we are able to express 
$\phi(\Lambda;x)$ by means of an interpolation determinant. 
For each $d=0,1,2,\ldots$, we denote by 
\begin{equation}
\bm_d(x)=\left(x^{\mu}\right)_{|\mu|=d}
=\left(x_1^{\mu_1}x_2^{\mu_2}x_3^{\mu_3}\right)_{\mu_1+\mu_2+\mu_3=d}
\end{equation}
the column vector of monomials of degree $d$ in $x=(x_1,x_2,x_3)$. 
Note that the number of such monomials is given by 
\begin{equation}
\dim_{\mathbb{K}}\mathbb{K}[x]_d=\binom{d+2}{2}=\dfrac{1}{2}(d+1)(d+2). 
\end{equation}
For each $k=0,1,\ldots,d$ we denote by 
\begin{equation}
\bm_d^{(k)}(x)
=\left(\partial_x^{\kappa}(x^\mu)\right)
_{|\mu|=d, |\kappa|=k}
=\left(
\partial_{x_1}^{\kappa_1}\partial_{x_2}^{\kappa_2}\partial_{x_3}^{\kappa_3}
(x_1^{\mu_1}x_2^{\mu_2}x_3^{\mu_3})\right)_{|\mu|=d, |\kappa|=k}
\end{equation}
the $\binom{d+2}{2}\times \binom{k+2}{2}$ matrix defined 
by the partial derivatives $\bm_d(x)$ of order $k$. 
(For $k<0$, we consider $\bm_d^{(k)}(x)$ as an empty matrix.)
Given an element 
\begin{equation}
\Lambda=de_0-\nu_1e_1-\ldots-\nu_9 e_9\in M_{3,9}, 
\end{equation}
we consider the homogeneous polynomial 
\begin{equation}\label{eq:FDet}
F(\Lambda;x)=\det\left(
\bm_{d}^{(\nu_1-1)}(p_1),
\ldots,\bm_{d}^{(\nu_9-1)}(p_9),\bm_{d}(x)\right) 
\end{equation}
of degree $d$ in $x=(x_1,x_2,x_3)$, 
where $\bm_{d}^{(k)}(p_j)$ stands for the matrix 
obtained from $\bm_{d}^{(k)}(x)$ by the substitution 
$x=x(p_j)$ as in \eqref{eq:xpj3}. 
Note here that 
\begin{equation}
\binom{d+2}{2}
-\dsum{j=1}{9}\binom{\nu_j+1}{2}-1
=-\dfrac{1}{2}\big(\ipr{c}{\Lambda}+\ipr{\Lambda}{\Lambda}\big) =0
\end{equation}
for any $\Lambda\in M_{3,9}$; 
this means that the number of column vectors in \eqref{eq:FDet} 
is equal to $\binom{d+2}{2}$. 
Also, 
from $\dim_{\mathbb{K}}L(\Lambda)=1$ it follows 
that $F(\Lambda;x)$ is a nonzero polynomial. 
Combining this fact with the normalization condition 
\eqref{eq:taunorm}, 
we obtain the following theorem.
\begin{theorem}\label{thm:det}
For each $\Lambda\in M_{3,9}$, 
define the homogeneous polynomial $F(\Lambda;x)$ in $x=(x_1,x_2,x_3)$ 
by the determinant \eqref{eq:FDet}. Then, 
the specialization of $F(\Lambda;x)$ to 
the canonical solution $x^C(t)$ is expressed in the form 
\begin{equation}
F(\Lambda;x^C(t))=C_\Lambda\,[\lambda-t]\,\dprod{j=1}{9}[\vep_j-t]^{\nu_j},
\qquad \lambda=\ipr{\Lambda}{\cdot}\in\mathfrak{h}_{3,9}^\ast,
\end{equation} 
with a nonzero constant $C_\Lambda\in\mathbb{K}$. With this 
normalization constant $C_{\Lambda}$, 
the homogeneous polynomial 
$\phi(\Lambda;x)\in L(\Lambda)$ is expressed as the determinant 
\begin{equation}
\phi(\Lambda;x)=C_{\Lambda}^{-1}F(\Lambda;x). 
\end{equation}
\end{theorem}

\par\medskip
Let us consider the action of translations $w=T_{\alpha}$ 
$(\alpha\in\Delta_{3,9}^{\re})$ on $\mathcal{R}$:
\begin{equation}
\arraycolsep=2pt
\begin{array}{rll}
T_{\alpha}(f_i)&=\tau^{T_{\alpha}.h_0-\deg(T_\alpha.h_0)h_0}\,
\dfrac{\phi(T_{\alpha}.(e_i+h_0);f)}
{\phi(T_{\alpha}.e_i;f)}\qquad&(i=1,2,3),\\[8pt]
T_{\alpha}(\tau_j)&=\tau^{T_{\alpha}.e_j-\deg(T_{\alpha}.e_j) h_0}\,
\phi(T_{\alpha}.e_j;f)\qquad&(j=1,\ldots,9). 
\end{array}
\end{equation}
These formulas are the elliptic difference Painlev\'e equations 
of type $(3,9)$ for the $f$ and $\tau$ variables. 
The polynomials $\phi(\Lambda;x)$ can be determined either 
recursively by \eqref{eq:phirec}, or by using the determinants 
of Theorem \ref{thm:det}. 
In this sense, these polynomials are computable in principle. 
%However, 
%the polynomial $\phi(T_{\alpha}(e_i);x)$ 
%and $\phi(T_{\alpha}(e_i+h_0);x)$ are too big to write down 
%explicitly.  
We will show later some explicit formulas for small $\phi(\Lambda;x)$. 

For $\alpha\in \Delta^{\re}_{3,9}$, the translation $T_{\alpha}$ 
is defined by 
\begin{equation} 
T_{\alpha}(\Lambda)=\Lambda-h+(1-\ipr{h}{\Lambda})c
\end{equation}
for any $\Lambda\in M_{3,9}$, where $h\in L_{3,9}$, 
$\alpha=\ipr{h}{\cdot}$. (Note that $\ipr{\alpha}{\alpha}=2$.)
As an example, we consider the case $T_{78}=T_{\alpha_{7}}$. 
(The argument below applies to any $T_{ij}=T_{\vep_{ij}}$ 
for mutually distinct $i,j\in\{4,5,6,7,8,9\}$). 
In this case, the discrete time evolution of the 
$f$ variables by $T_{78}$ 
can be written in the form
\begin{equation} 
T_{78}(f_i)=\dfrac{G_i(f_1,f_2,f_3)}{F_i(f_1,f_2,f_3)} \quad(i=1,2,3)
\end{equation}
where $F_i(x)=\phi(T_{78}(e_i);x)$ and 
$G_i(x)=\phi(T_{78}(e_i+h_0);x)$. 
For $i=1$ we have 
\begin{equation}
T_{78}(e_1)%=e_i-h_8+c 
=3e_0-e_2-e_3-e_4-e_5-e_6-2e_7-e_9. 
\end{equation} 
This means that $F_1(x)=\phi(T_{78}(e_1);x)$ is a homogeneous 
polynomial of degree $3$ with multiplicities of zeros 
$(0,1,1)$ at $(p_1,p_2,p_3)$, and $(1,1,1,2,0,1)$ at 
$(p_4,p_5,p_6,p_7,p_8,p_9)$. 
In particular $F_1(x)$ can be written in the form 
\begin{equation}
F_1(x)=a_{0} x_1^3+ a_1 x_1^2 x_2+a_{2} x_1^2 x_3+
a_{3} x_1 x_2^2 +a_{4}x_1x_2x_3+a_{5} x_1x_3^2+
a_{6} x_2^2 x_3+a_{7} x_2 x_3^2. 
\end{equation}
Observe that the monomials $x_2^3$ and $x_3^3$ are missing 
in this formula;
this is because $F_1(x)$ should have zeros at $p_2=(0:1:0)$ 
and $p_3=(0:0:1)$.
(The coefficients $a_k$ could be determined in principle 
from the pattern of multiplicities of zeros at $(p_4,\ldots,p_9)$.)
Similarly from 
\begin{equation}
T_{78}(e_1+h_0)
=4e_0-e_1-2e_2-2e_3-e_4-e_5-e_6-2e_7-e_9,
\end{equation}
we see that $G_i(x)$ is of degree 4. 
Since $s_0 T_{78}(e_1)=T_{78}(e_1+h_0)$, 
we know that $G_1(x)$ is in fact determined from $F_1(x)$ as 
$G_1(x)=x_1^3x_2^2x_3^2\,{}^{s_0}\!F(x^{-1})$. 
These polynomials $F_i(x)$ and $G_i(x)$ 
are in fact too big to write down explicitly. 
A way to {\em see} this time evolution of the $f$ variables is 
to decompose $T_{78}$ into two steps by using the expression 
\begin{equation}
T_{78}=w^2,\qquad w=s_{11'7}s_{22'7}s_{33'7}s_{89}s_{78},
\end{equation}
where $\{1',2',3'\}=\{4,5,6\}$. 
If we set $g_i=w(f_i)$, we obtain
\begin{equation}\label{eq:fgTf}
g_i=\dfrac{Q_i(f_1,f_2,f_3)}{P_i(f_1,f_2,f_3)},\quad
T_{78}(f_i)=\dfrac{S_i(g_1,g_2,g_3)}{R_i(g_1,g_2,g_3)}
\quad(i=1,2,3), 
\end{equation} 
where 
$P_i(x)=\phi(w(e_i);x)$ and $Q_i(x)=\phi(w(e_i+h_0);x)$; 
$R_i(x)$ and $S_i(x)$ are determined as 
$R_i(x)={}^{w}\!P_i(x)$ and $S_i(x)={}^{w}\!Q_i(x)$,
by applying $w$ to the coefficients. 
Since
\begin{equation}
w(e_i)=e_0-e_{j'}-e_{k'},\quad
w(e_i+h_0)=2e_0-e_1-e_2-e_3-e_{j'}-e_{k'}
\end{equation} 
for $\{i,j,k\}=\{1,2,3\}$, 
we see that the corresponding $\phi(\Lambda;x)$ have 
degree 1 and 2, respectively. 
In fact we have
\begin{equation}
\phi(e_0-e_a-e_b;x)
=
\dfrac{[\vep_{1a}][\vep_{1b}][\vep_{1ab}]}
{[\vep_{12}][\vep_{13}][\vep_{123}]}x_1
-\dfrac{[\vep_{2a}][\vep_{2b}][\vep_{2ab}]}
{[\vep_{12}][\vep_{23}][\vep_{123}]}x_2
+\dfrac{[\vep_{3a}][\vep_{3b}][\vep_{3ab}]}
{[\vep_{13}][\vep_{23}][\vep_{123}]}x_3
\end{equation}
for $1\le a<b\le 9$, and 
\begin{equation}
\begin{array}{ll}
\phi(2e_0-e_1-e_2-e_3-e_a-e_b;x)\\[4pt]
=
-\dfrac{[\vep_{23a}][\vep_{23b}][\vep_{1ab}]}
{[\vep_{12}][\vep_{13}][\vep_{123}]}x_2x_3
+\dfrac{[\vep_{13a}][\vep_{13b}][\vep_{2ab}]}
{[\vep_{12}][\vep_{23}][\vep_{123}]}x_1x_3
-\dfrac{[\vep_{12a}][\vep_{12b}][\vep_{3ab}]}
{[\vep_{13}][\vep_{23}][\vep_{123}]}x_1x_2
\end{array}
\end{equation}
for $4\le a<b\le 9$. 
%for any $a,b\in\{4,5,6,7,8,9\}$ with $a\ne b$. 
Hence we obtain the explicit formulas for $P_i.Q_i,R_i,S_i$: 
\begin{equation}\label{eq:PQ}
\arraycolsep=2pt
\begin{array}{rl}
P_i(x)&=
\dfrac{[\vep_{1i'}][\vep_{17}][\vep_{1i'7}]}
{[\vep_{12}][\vep_{13}][\vep_{123}]}x_1
-\dfrac{[\vep_{2i'}][\vep_{27}][\vep_{2i'7}]}
{[\vep_{12}][\vep_{23}][\vep_{123}]}x_2
+\dfrac{[\vep_{3i'}][\vep_{37}][\vep_{3i'7}]}
{[\vep_{13}][\vep_{23}][\vep_{123}]}x_3,
\\[10pt]
Q_i(x)&=
-\dfrac{[\vep_{23i'}][\vep_{237}][\vep_{1i'7}]}
{[\vep_{12}][\vep_{13}][\vep_{123}]}x_2x_3
+\dfrac{[\vep_{13i'}][\vep_{137}][\vep_{2i'7}]}
{[\vep_{12}][\vep_{23}][\vep_{123}]}x_1x_3
-\dfrac{[\vep_{12i'}][\vep_{127}][\vep_{3i'7}]}
{[\vep_{13}][\vep_{23}][\vep_{123}]}x_1x_2,
\\[10pt]
R_i(x)&=
-\dfrac{[\vep_{i1'}][\vep_{1'79}][\vep_{i1'8}^{-}]}
{[\vep_{1'2'}][\vep_{1'3'}][\vep_{123}]}x_1
+\dfrac{[\vep_{i2'}][\vep_{2'79}][\vep_{i2'8}^{-}]}
{[\vep_{1'2'}][\vep_{2'3'}][\vep_{123}]}x_2
-\dfrac{[\vep_{i3'}][\vep_{3'79}][\vep_{i3'8}^{-}]}
{[\vep_{1'3'}][\vep_{2'3'}][\vep_{123}]}x_3,
\\[10pt]
S_i(x)&=
-\dfrac{[\vep_{jk1'}][\vep_{2'3'8}^{-}][\vep_{i1'8}^{-}]}
{[\vep_{1'2'}][\vep_{1'3'}][\vep_{123}]}x_2x_3
+\dfrac{[\vep_{jk2'}][\vep_{1'3'8}^{-}][\vep_{i2'8}^{-}]}
{[\vep_{1'2'}][\vep_{2'3'}][\vep_{123}]}x_1x_3
-\dfrac{[\vep_{jk3'}][\vep_{1'2'8}^{-}][\vep_{i3'8}^{-}]}
{[\vep_{1'3'}][\vep_{2'3'}][\vep_{123}]}x_1x_2
\end{array}
\end{equation}
for $\{i,j,k\}=\{1,2,3\}$, where 
$\vep_{ij8}^{-}=\vep_{ij8}-\delta$. 

We remark that the translation $T_{\alpha}$ for 
any $\alpha\in\Delta^{\re}_{3,9}$ can be 
expressed as $T_{\alpha}=v\,T_{78}\,v^{-1}$ for 
some $v\in W_{3,9}$ such that $v(\alpha_7)=\alpha$. 
If we introduce the dependent variables 
$\varphi_i=v(f_i)$, $\psi_i=v(g_i)$ ($i=1,2,3$), 
the discrete time evolution of these 
variables by $T_{\alpha}$ 
can be expressed in the same form as in the case of $T_{78}$. 

Explicit description for the time evolutions of 
the elliptic difference Painlev\'e equation  
is discussed also in \cite{ORG2} and \cite{Murata} from different 
viewpoints. 

\section{In terms of geometry of plane curves}
\label{planecurves}

The discrete time evolution of type $T_{ij}$ ($i,j\in\{1,\ldots,9\}; i\ne j$) 
for the elliptic difference Painlev\'e equation can be described 
by means of geometry of plane cubic curves. In this final section 
we give an explanation of this fact in the scope of this paper. 
\par\medskip
Take three constants $c_1,c_2,c_3\in\mathbb{C}$ such that 
\begin{equation}
[c_1+c_2+c_3]\ne 0,\qquad [c_i-c_j]\ne 0\quad(1\le i<j\le 3),
\end{equation}
and set $c_0=-c_1-c_2-c_3$. With these constants fixed, 
let us consider 
the holomorphic mapping $p: \mathbb{C}\to \mathbb{P}^2(\mathbb{C})$ 
defined by 
\begin{equation}\label{eq:parC0}
\begin{array}{c}
p(u)=(x_1(u):x_2(u):x_3(u)) \quad(u\in\mathbb{C}),\\
x_i(u)=[c_0+c_i-u][c_j-u][c_k-u]\qquad(\{i,j,k\}=\{1,2,3\}).
\end{array}
\end{equation}
(By the quasi-periodicity of the function $[u]$, 
this mapping induces a holomorphic mapping 
$\overline{p}: E=\mathbb{C}/\Omega \to \mathbb{P}^2(\mathbb{C})$ as well.) 
We denote by $C_0=\overline{p(\mathbb{C})}$ the plane curve 
obtained by the parametrization \eqref{eq:parC0}.
The defining equation for this curve $C_0$ is given explicitly by
\comment{
\begin{equation}
\begin{array}{ll}
\left(2\dfrac{[c_0][c_0+c_2-c_3]'}{[0]'[c_2-c_3]}x_1
\!-\!\dfrac{[c_0+c_1-c_2]}{[c_1-c_2]}x_2
\!+\!\dfrac{[c_0+c_1-c_3]}{[c_1-c_3]}x_3\right)x_2x_3\\[8pt]
\mbox{}+
\left(\dfrac{[c_0+c_2-c_1]}{[c_2-c_1]}x_1
\!+\!2\dfrac{[c_0][c_0+c_3-c_1]'}{[0]'[c_3-c_1]}x_2
\!-\!\dfrac{[c_0+c_2-c_3]}{[c_2-c_3]}x_3
\right)x_3x_1\\[8pt]
\mbox{}+
\left(-\dfrac{[c_0+c_3-c_1]}{[c_3-c_1]}x_1
\!+\!\dfrac{[c_0+c_3-c_2]}{[c_3-c_2]}x_2
\!+\!2\dfrac{[c_0][c_0+c_1-c_2]'}{[0]'[c_1-c_2]}x_3
\right)x_1x_2=0,
\end{array}
\end{equation}
or
}
\begin{equation}
\begin{array}{ll}
-\dfrac{[c_0+c_3-c_1]}{[c_3-c_1]}x_1^2x_2
+\dfrac{[c_0+c_2-c_1]}{[c_2-c_1]}x_1^2x_3
+
\dfrac{[c_0+c_3-c_2]}{[c_3-c_2]}x_1x_2^2
\\[10pt]
\mbox{}+
2\dfrac{[c_0]}{[0]'}
\left(\dfrac{[c_0+c_2-c_3]'}{[c_2-c_3]}\!+\!
\dfrac{[c_0+c_3-c_1]'}{[c_3-c_1]}\!+\!
\dfrac{[c_0+c_1-c_2]'}{[c_1-c_2]}
\right)x_1x_2x_3
\\[10pt]
\mbox{}-
\dfrac{[c_0+c_2-c_3]}{[c_2-c_3]}x_1x_3^2
-\dfrac{[c_0+c_1-c_2]}{[c_1-c_2]}x_2^2x_3
+\dfrac{[c_0+c_1-c_3]}{[c_1-c_3]}x_2x_3^2=0,
\end{array}
\end{equation}
where $[u]'$ stands for the derivative of $[u]$. 
(The coefficient of $x_1x_2x_3$ can be written in various ways.)
We remark that this definition of $p$ and $C_0$ is related to 
that of $p_{\lambda,\mu}$ and $C_{\lambda,\mu}$ 
for the case $m=3$ in Section \ref{linearization} by the 
change of variables 
\begin{equation}
\lambda=c_0, \quad
\mu_i=c_i+\dfrac{\vep_0}{3}\quad(i=1,2,3),\quad t=u+\dfrac{\vep_0}{3}.
\end{equation}
In particular, 
the $W_{3,n}$-equivariant meromorphic mapping 
$\varphi_{3,n}: \mathfrak{h}_{3,n} \ratto \mathbb{X}_{3,n}$  
can be realized by means of point configurations on one single 
curve $C_0\subset\mathbb{P}^2(\mathbb{C})$: 
\begin{equation}
\varphi_{3,n}(\vep)=[p(u_1),\cdots,p(u_n)],\qquad u_j=\vep_j-\dfrac{\vep_0}{3}
\quad(j=1,\ldots,n) 
\end{equation}
for each generic $\vep=(\vep_0,\vep_1,\ldots,\vep_n)\in\mathfrak{h}_{3,n}$. 
%(The coordinate $u$ for $C_0$ is related to that for 
%$C_{\lambda,\mu}$ through the formula $u=t-\dfrac{\vep_0}{3}$.) 
Thanks to the condition $c_0+c_1+c_2+c_3=0$, 
we see that a set of $3d$ points $p(a_1),\ldots p(a_{3d})$ on $C_0$ 
is realized as the intersection of $C_0$ and a curve of degree $d$ 
if and only if $[a_1+\cdots+a_{3d}]=0$.  
In particular, 
three points $p(a_1), p(a_2), p(a_3)$ on $C_0$ are colinear 
if and only if $[a_1+a_2+a_3]=0$. 
The affine Weyl group $W_{3,n}$ acts on these variables 
$u_1,\ldots,u_n$ as follows:
\begin{equation}\label{eq:Wonu}
\begin{array}{ll}
s_0(u_j)=\left\{
\begin{array}{ll}
u_j-\dfrac{2}{3}(u_1+u_2+u_3)\quad&(j=1,2,3),\\[6pt]
u_j+\dfrac{1}{3}(u_1+u_2+u_3)\quad&(j=4,\ldots,n).
\end{array}\right.\\[6pt]
s_k(u_j)=u_{s_k(j)}\qquad(k=1,\ldots,n-1; j=1,\ldots,n). 
\end{array}
\end{equation} 

Before going further, we remark that any irreducible cubic curve 
in $\mathbb{P}^{2}(\mathbb{C})$ can be obtained by a projective linear
transformation from a curve of the form $C_0$ 
with $[u]$ and $c_1,c_2,c_3$ appropriately chosen. 
In fact the curve $C_0$ is related 
with the Weierstrass canonical form of a cubic curve
in the following way. 
Consider the case when $\mbox{rank}\,\Omega=2$, and 
for $[u]$ take 
the Weierstrass sigma function $\sigma(u)=\sigma(u;\Omega)$ 
associated with the period lattice 
$\Omega=\mathbb{Z}\omega_1\oplus \mathbb{Z}\omega_2$. 
If we set 
\begin{equation}
c_0=-\dfrac{\omega_1+\omega_2}{2},\quad
c_1=\dfrac{\omega_1}{2},\quad 
c_2=\dfrac{\omega_2}{2},\quad c_3=0,
\end{equation}
then the parametrization of $C_0$ is given by 
\begin{equation}
\begin{array}{ll}
x_1=\sigma(u)\sigma(c_2+u)\sigma(c_2-u)
=\sigma(c_2)^2\sigma(u)^3(\wp(u)-\wp(c_2)),\\[2pt]
x_2=\sigma(u)\sigma(c_1+u)\sigma(c_1-u)
=\sigma(c_1)^2\sigma(u)^3(\wp(u)-\wp(c_1)),\\[2pt]
x_3=\sigma(c_0-u)\sigma(c_1-u)\sigma(c_2-u)
=-\frac{1}{2}\sigma(c_0)\sigma(c_1)\sigma(c_2)\sigma(u)^3\wp'(u),
\end{array}
\end{equation}
where $\wp(u)=\wp(u;\Omega)$ is the Weierstrass $\wp$ function 
associated with $\Omega$. 
Hence the curve $C_0$ is transformed into the canonical form 
\begin{equation}
y_1y_3^2=4(y_2-\wp(c_0)y_1)(y_2-\wp(c_1)y_1)(y_2-\wp(c_2)y_1)
\end{equation}
by the projective linear transformation
\begin{equation}
\begin{array}{lll}
x_1=\sigma(c_2)^2(y_2-\wp(c_2)y_1),\qquad
x_2=\sigma(c_1)^2(y_2-\wp(c_1)y_1),\\[2pt]
x_3=-\frac{1}{2}\sigma(c_0)\sigma(c_1)\sigma(c_2)y_3. 
\end{array}
\end{equation}
This implies that any smooth cubic curve can be expressed in 
the form of $C_0$ up to a projective linear transformation. 
Note that through this transformation to the Weierstrass canonical 
form, formula \eqref{eq:Wonu} recovers the same Weyl group action 
in the parametrization by the $\wp$ function as in \cite{Sakai}. 

\par\medskip 
In what follows, we consider the translation $T_{89}\in W_{3,9}$ 
as an example, and describe the corresponding Cremona 
transformation in the language of geometry of plane curves.  
Namely, given an generic configuration 
$[p_1,\ldots,p_9,q]\in\mathbb{X}_{3,10}$, 
we explain in geometric terms how to 
specify $\overline{p}_1,\ldots,\overline{p}_9$ and 
$\overline{q}$ in $\mathbb{P}^2(\mathbb{C})$ such that  
\begin{equation} 
[p_1,\ldots,p_9,q].T_{89}=
[\overline{p}_1,\ldots,\overline{p}_9,\overline{q}]. 
\end{equation}

We first consider the case where all the 10 points $p_1,\ldots,p_9,q=p_{10}$ 
are on a smooth cubic curve $C$. 
Given four points $p,q,p',q'\in C$, 
we say that $p+q=p'+q'$ {\em under the addition of $C$} if 
the third intersection point of the 
line $L_{p,q}$, passing through $p,q$, with $C$ coincides 
with that of the line $L_{p',q'}$.
\begin{lemma}\label{lem:onC}
Let $[p_1,\ldots,p_9,p_{10}]\in \mathbb{X}_{3,10}$ be 
generic 
and assume that the 10 points $p_1,\ldots,$ $p_9,p_{10}$ 
are on a smooth cubic curve $C$.
Then the action of $T_{89}$ on $[p_1,\ldots,p_9,p_{10}]$ 
is expressed in the form 
\begin{equation} 
[p_1,\ldots,p_7,p_8,p_9,p_{10}].T_{89}
=[p_1,\ldots,p_7,\overline{p}_8,\overline{p}_9,\overline{p}_{10}]
\end{equation}
by using the three points 
$\overline{p}_8,\overline{p}_9,\overline{p}_{10}\in C$ that are 
determined by the following three conditions. 
\newline 
\quad$(1)$\quad The $9$ points $p_1,\ldots,p_8,\overline{p}_{9}$ form 
the base points of a pencil of cubic curves.  
\newline 
\quad$(2)$\quad $\overline{p}_8+\overline{p}_9=p_8+p_9$ under the addition 
of $C$.
\newline 
\quad$(3)$\quad $\overline{p}_{9}+\overline{p}_{10}=p_8+p_{10}$ 
under the addition of $C$. 
\end{lemma}
In order to prove Lemma \ref{lem:onC}, 
by a projective linear transformation, we may assume that this curve $C$ 
is of the form $C_0$, and that the 10 points are parametrized as  
\begin{equation} 
[p_1,\ldots,p_9,p_{10}]
=[p(u_1),\ldots,p(u_9),p(u_{10})]. 
\end{equation}
As we already know, such a configuration is transformed by $T_{89}$ 
into 
\begin{equation}
\begin{array}{l}
[p_1,\ldots,p_9,p_{10}].T_{89}=
[\overline{p}_1,\ldots,\overline{p}_9,\overline{p}_{10}],
\\[2pt]
\overline{p}_j=p(\overline{u}_j),
\quad \overline{u}_j=T_{89}(u_j)\qquad(j=1,\ldots,10). 
\end{array}
\end{equation}
Since  
\begin{equation}
\begin{array}{l}
T_{89}(u_j)=u_j\qquad(j=1,\ldots,7),\\ 
T_{89}(u_8)=u_8-\delta,\quad
T_{89}(u_9)=u_9+\delta,\\
T_{89}(u_{10})=u_{10}+u_{8}-u_{9}-\delta,\\
u_1+\cdots+u_9=-\delta, 
\end{array}
\end{equation}
the new coordinates 
$\overline{u}_j$ ($j=1,\ldots,10$) 
are determined by the conditions 
\begin{equation}
\begin{array}{lll}
(0)&\overline{u}_j=u_j\qquad(j=1,\ldots,7),\\
(1)& u_1+\cdots+u_8+\overline{u}_9=0,\\
(2)&\overline{u}_8+\overline{u}_9=u_8+u_9,\\
(3)&\overline{u}_9+\overline{u}_{10}=u_{8}+u_{10}.
\end{array} 
\end{equation} 
Lemma \ref{lem:onC} is a paraphrase of 
this characterization of $\overline{u}_j$ $(j=1,\ldots,10)$ 
in geometric terms. 
We remark that the point $\overline{p}_9$ is 
determined only from $p_1,\ldots,p_8$, and does not depend on 
the position of $p_9$, 
while $\overline{p}_8$ depends essentially on $p_9$. 

Lemma \ref{lem:onC} can be extended to the general case as follows. 
\begin{theorem}\label{thm:geom}
Let $[p_1,\ldots,p_9,q]$ be a configuration of 10 points 
in $\mathbb{P}^2(\mathbb{C})$ in general position. 
Suppose that this configuration is generic, and take   
two smooth cubic curves $C_0$ and $C$ such that 
\begin{equation} 
p_1,\ldots,p_8,p_9\in C_0,\quad\mbox{and}\quad 
p_1,\ldots,p_8,q\in C,
\end{equation}
respectively. 
Then the action of the translation $T_{89}$ on the configuration 
$[p_1,\ldots,p_9,q]\in\mathbb{X}_{3,10}$ is expressed as 
\begin{equation}
[p_1,\ldots,p_9,q].T_{89}=[p_1,\ldots,p_7,\overline{p}_8,\overline{p}_{9},
\overline{q}],
\end{equation}
in terms of the points $\overline{p}_8,\overline{p}_9$ 
on $C_0$ and $\overline{q}\in C$ that are 
determined by the following conditions\,$:$  
\newline
\quad$(1)$\ \ The 9 points $p_1,\ldots,p_{8},\overline{p}_{9}\in C_0$ 
form the base points of a pencil of cubic curves.
\newline
\quad$(2)$\ \ Under the addition of $C_0$, $\overline{p}_8+\overline{p}_9=
p_8+p_9$. 
\newline
\quad$(3)$\ \ Under the addition of $C$, $\overline{p}_9+\overline{q}=
p_8+q$. 
\newline
In particular $\overline{p}_9$ is determined as the ninth point 
in the intersection of $C_0$ and $C$. 
\end{theorem} 
From Lemma \ref{lem:onC} applied to $C_0$, we have
\begin{equation}
[p_1,\ldots,p_7,p_8,p_9].T_{89}=
[p_1,\ldots,p_7,\overline{p}_8,\overline{p}_9]
\end{equation}
with $\overline{p}_8, \overline{p}_9\in C_0$ determined by 
the conditions (1), (2) of Theorem \ref{thm:geom}. (This part 
does not depend on the tenth point.) 
Hence, the action of $T_{89}$ on $[p_1,\ldots,p_8,p_9,q]$ 
can be written as
\begin{equation}\label{eq:TA}
[p_1,\ldots,p_7,p_8,p_9,q].T_{89}=
[p_1,\ldots,p_7,\overline{p}_8,\overline{p}_9,\overline{q}]
\end{equation}
for some $\overline{q}\in\mathbb{P}^{2}(\mathbb{C})$. 
We remark here that $T_{89}$ can be expressed in the form 
\begin{equation}
T_{89}=w\,s_{89},\qquad w=s_{128}s_{348}s_{567}s_{348}s_{128}\in 
W_{3,8},
\end{equation} 
where $W_{3,8}=\br{s_0,s_1,\ldots,s_7}$. 
Hence, by applying $s_{89}\,s_{9,10}\in W_{3,10}$ 
to \eqref{eq:TA} from the right, 
we obtain
\begin{equation} \label{eq:TB}
[p_1,\ldots,p_7,p_8,q,p_9].w=
[p_1,\ldots,p_7,\overline{p}_9,
\overline{q},\overline{p}_8]
\qquad \mbox{in}\quad \mathbb{X}_{3,10}. 
\end{equation}
(Note that $s_{9,10}$ commute with $w\in W_{3,8}$.)
Since $w\in W_{3,8}$, this formula projects to  
\begin{equation} \label{eq:TC}
[p_1,\ldots,p_7,p_8,q].w=
[p_1,\ldots,p_7,\overline{p}_9,\overline{q}]
\qquad \mbox{in}\quad \mathbb{X}_{3,9}. 
\end{equation} 
This implies that $\overline{q}$ does not depend on $p_9$. 
(This fact can be seen clearly in formula \eqref{eq:PQ} 
for $T_{78}$. In fact, none of 
the polynomials $P_i, Q_i, R_i, S_i$ depends on 
the parameter $\vep_8$.)
Hence, by considering the configuration 
$[p_1,\ldots,p_8,\overline{p}_9,q]$ on $C$ with 
$\overline{p}_9$ replaced for $p_9$, 
we have 
\begin{equation}
[p_1,\ldots,p_7,p_8,\overline{p}_9,q].T_{89}
=[p_1,\ldots,p_7,p_8,\overline{p}_9,\overline{q}]. 
\end{equation}
(The 8th and 9th components remain invariant since 
$p_1,\ldots,p_8,\overline{p}_9$ are already the base points 
of a pencil of cubic curves containing $C$.)
Then by applying Lemma \ref{lem:onC} to $C$, we 
conclude that $\overline{q}$ is determined by condition (3). 
\par\medskip
Geometric description of discrete time evolutions of type $T_{ij}$ as 
described above is proved in \cite{KMNOY1} by a more geometric 
argument based on the results of \cite{Manin} and \cite{Nagata}. 
We remark that 
this geometric approach has been employed 
in the study of hypergeometric solutions to elliptic 
and multiplicative discrete Painlev\'e equations 
in \cite{KMNOY1}, \cite{KMNOY2}. 
It is also used by \cite{Tsuda} in order to clarify the relationship 
between the elliptic difference Painlev\'e equation and  
the integrable mapping of Quispel-Roberts-Thompson \cite{QRT}. 

%%%%%%%%%%%%%%%%%%%%%%%%%%%%%%%%%%%%%%%%%%%%%%%%%%%%%%%%%%%%%%%%%%%%

\end{document}